\documentstyle[prb,preprint,aps]{revtex}
\begin{document}
\draft
\title{A Biased Resistor Network Model for Electromigration Failure
\\ and Related Phenomena in Metallic Lines }

\author{C. Pennetta$^{1,2}$, E. Alfinito$^{1,2}$, L. Reggiani$^{1,2}$, 
\\ F. Fantini$^{3}$, I. DeMunari$^{4}$, A. Scorzoni$^{5}$}

\address{$^1$ INFM - National Nanotechnology Laboratory, 
Via Arnesano, I-73100, Lecce, Italy,
\\$^2$ Dipartimento di Ingegneria dell'Innovazione,
Universit\`a di Lecce, Italy, \\
$^3$ INFM and Dipartimento di Ingegneria dell'Informazione, 
Universit\`a di Modena,\\ Via Vignolese 905, I-41100 Modena, Italy,\\
$^4$ INFM and Dipartimento di Ingegneria dell'Informazione, 
Universit\`a di Parma, \\Parco Area delle Scienze 181/A, I-43100  Parma, 
Italy,\\
$^5$ INFN and Dipartimento di Ingegneria Elettronica e dell'Informazione, 
Universit\`a di Perugia,\\ Via G. Duranti 93, I-06125 Perugia, Italy.
\thanks{Corresponding authors e-mail: cecilia.pennetta@unile.it }}
\date{\today}
\maketitle
\begin{abstract}
Electromigration phenomena in metallic lines are studied by using a biased 
resistor network model. The void formation induced by the electron wind 
is simulated by a stochastic process of resistor breaking, while the growth of
mechanical stress inside the line is described by an antagonist process of 
recovery of the broken resistors. The model accounts for the existence of 
temperature gradients due to current crowding and Joule heating. Alloying 
effects are also accounted for. Monte Carlo simulations allow the study
within a unified theoretical framework of a variety of relevant features
related to the electromigration. The predictions of the model are 
in excellent agreement with the experiments and in particular with the 
degradation towards electrical breakdown of stressed Al-Cu thin metallic 
lines. Detailed investigations refer to the damage pattern, the distribution 
of the times to failure (TTFs), the generalized Black's law, the time 
evolution of the resistance, including the early-stage change due to alloying 
effects and the electromigration saturation appearing at low current densities
or for short line lengths. The dependence of the TTFs on the length and  
width of the metallic line is also well reproduced. Finally, the model 
successfully describes the resistance noise properties under steady state 
conditions. 

\end{abstract}
\vspace{1.5cm}
\pacs{PACS numbers: 66.30.Qa; 85.40.Qx; 85.40.Ls; 64.60.Ak}
\narrowtext
\section{Introduction}
The phenomenon of electromigration (EM) is typical of metallic conductors and 
consists in a non-steady atomic transport driven by electronic currents of 
high density \cite{ohring,em_review}. The non-steady atomic transport gives 
rise to the formation and growth of voids and hillocks in different regions of
the conductor (EM damage) \cite{ohring,em_review}. This damage cumulates 
progressively and, when the current is applied for a sufficiently long time, 
a void grows enough to break completely the metallic line implying an 
irreversible failure process. The time required for this process defines the
time to failure (TTF) of the metallic line \cite{ohring,em_review} 
(though alternative failure criteria can be found in the literature 
\cite{ohring,em_review,nat_IC}). The importance of EM is largely due to the 
fact that it is the most common mechanism of failure of the metallic 
interconnects present in any electronic device \cite{ohring,em_review,nat_IC}.
As a consequence, a huge number of experimental 
\cite{cho,healing,koch,filippi_jap78,filippi_apl66,filippi_96_02,scorz_j96,scorzoni96,foley98,jones,capasso,pattern_1,pattern_2} and theoretical 
\cite{shatzkes86,lloyd91,korhonen93,bradley,knowlton97,krug98,tammaro99,proost,sasagawa} studies have been and are yet devoted to the subject, especially
in the context of modern nanoelectronics. 

The central issue in EM degradation phenomena is the determination of   
the TTF and its statistical properties \cite{ohring,em_review,nat_IC}.
TTFs are measured under very high stress conditions (currents and temperature 
much higher than those corresponding to the usual operating conditions of the 
devices) in the so called accelerated tests \cite{ohring,em_review}. 
The extrapolation to normal operating conditions is generally performed on the
basis of three frequently adopted assumptions \cite{ohring,em_review}. First, 
TTFs are taken to follow a lognormal distribution (which means that the 
logarithms of TTFs are normally distributed). This distribution is then 
characterized by two parameters: the median time to failure, $t_{50}$, and the
shape factor, $s$, where $t_{50}$ is the time corresponding to the failure of 
50\% of the lines in the statistical sample and $s$ is the lognormal 
root-mean-square deviation \cite{ohring}. This distribution has been observed 
in many EM failure tests 
\cite{ohring,em_review,nat_IC,filippi_jap78,filippi_apl66,capasso} and 
recently, new testing techniques \cite{capasso} allowing the analysis of very 
large statistical samples, have shown that the TTF distribution follows a 
perfect lognormal behavior down to four shape factors \cite{capasso}. In spite 
of this evidence, no satisfactory eplanation has been given until now for
the lognormality of the TTF distribution \cite{ohring,em_review,lloyd91}.

The second assumption concerns the independence of the distribution shape 
factor of the stress conditions. Actually, $s$ was found to be independent of 
temperature in all the range of values usually considered in accelerated tests
\cite{ohring,em_review,nat_IC,filippi_jap78,filippi_apl66,capasso}. On the 
contrary, a broadening of the distribution has been observed at the lowest 
current densities used in these tests 
\cite{nat_IC,filippi_jap78,filippi_apl66}. This broadening of the distribution 
at low stress conditions has crucial implications on the evaluation of the 
minimum time to failure, i.e. the time corresponding to the first failure of a
line of the family \cite{filippi_jap78,filippi_apl66}. We will discuss this 
point in Sec. III.B. together with the results of our simulations. 

The third assumption concerns the validity of the following empirical law, 
known as Black's law \cite{ohring,em_review,black67}, relating $t_{50}$ to the
current density, $j$, and temperature, $T$:
\begin{equation} 
t_{50} = C j^{-n} \exp \Bigr[{E \over k_B T} \Bigl]
\label{eq:Black}
\end{equation}
where $C$ is a fitting amplitude, $n$ the so called current exponent, $E$ the 
EM activation energy and $k_B$ the Boltzmann constant. The validity of this 
law is confirmed by many experiments \cite{ohring,em_review,nat_IC}. However, 
the simple power law behavior described by Eq.~(\ref{eq:Black}), generally 
holds in a limited range of intermediate current densities 
\cite{ohring,em_review,nat_IC}. At high current densities, another power law 
with a different (higher) exponent is frequently observed and usually  
attributed to Joule heating effects \cite{ohring,em_review,nat_IC,tammaro99}, 
while at low current densities $t_{50}$ deviates from a power law, showing 
a tendency to diverge \cite{em_review,filippi_jap78}. Furthermore, even in the 
intermediate range of $j$ values, there is no agreement in the literature about
the value of the current exponent 
\cite{ohring,em_review,nat_IC,shatzkes86,tammaro99}. We will come back 
on this subject in Sec. III.B, discussing our results.

The geometry of the line plays a crucial role on the EM damage 
\cite{ohring,em_review,nat_IC,filippi_96_02,blech}. In fact, the depletion and
accumulation of mass in different regions of the film, under the driving force
exerted by the electronic current, determines the growth of mechanical stress 
gradients. The last ones give rise to an atomic back-flow which contrasts the 
EM process \cite{ohring,em_review,blech}. As the strength of these gradients 
depends strongly on the line geometry 
\cite{ohring,em_review,filippi_96_02,blech}, geometrical parameters have a 
fundamental role in the occurrence of failures. In particular, the competition
between the driving force of the electronic current and the action of 
mechanical stress gradients, results in the existence of a current density 
threshold below which EM is stopped \cite{blech}. This condition was expressed
by Blech and Herring \cite{blech} in the following equation which relates the 
threshold current density with the length of the line:
\begin{equation}
(jL)_{c}= {(\Omega_a \Delta \sigma) \over (\rho Z^*e)}
\label{eq:Blech}
\end{equation}
where $\Omega_a$ is the atomic volume, $\Delta \sigma$ is the maximum value of
the mechanical stress difference between the line terminals that a line of 
length $L$ can bear, $\rho$ is the resistivity of the line, $e$ the electronic
charge and $Z^*$ an average effective charge \cite{ohring,em_review,blech}. 
Equation ~(\ref{eq:Blech}), known as Blech's law  \cite{blech}, defines the 
so called threshold product, $(jL)_{c}$. We will discuss the role of the line 
geometry on the EM failure process in Sec. III.C.

Another fundamental ingredient in the understanding of the EM damage of 
interconnects is represented by the granular structure of the materials 
employed, Al, Cu, Ag, Al alloys, etc \cite{ohring,em_review,nat_IC}. 
Furthermore, it must be noted that a high degree of disorder is usually 
present in alloy films (typically Al-Cu, Al-Si) due to alloying effects 
\cite{em_review,scorz_j96,scorzoni96,pen_physd,dekker,kao} 
and to thermal gradients \cite{ohring,em_review}.

A large number of models has been proposed for the study of EM 
\cite{shatzkes86,lloyd91,korhonen93,bradley,knowlton97,krug98,tammaro99,proost,sasagawa}.
Many of them are microscopic models which address the problem of identifying 
the mechanisms responsible for the degradation process in terms of the 
peculiarities of the material considered 
\cite{korhonen93,knowlton97,krug98,proost,sasagawa}. Then, by using 
appropriate kinetic equations, some specific features of the damage process 
are determined and compared with experiments 
\cite{korhonen93,knowlton97,krug98,proost,sasagawa}. If the peculiarities of 
the material are sufficiently well accounted for, the predictivity of these 
models is very high \cite{korhonen93,knowlton97,sasagawa} and therefore they 
can be very useful for applicative purposes. However, the approach used by 
many of these models intrinsically limits their predictivity to some specific 
features of the EM damage. 

Another class of EM models has also been developed in the literature
\cite{bradley,pen_physd}, based on a ``coarse grain'' random resistor network 
(RN) approach \cite{chakrabarti,stauffer,sahim98,torquato}. Actually, the use 
of this kind of models is particularly appropriate in consequence of the
granular structure of the materials used for the interconnects. Indeed, it has
been observed that the atomic transport through grain boundaries and 
interfaces (transport channels) far exceeds that through the bulk of the 
grains \cite{ohring,em_review}. Therefore, it is generally possible to neglect
mass transport everywhere except within these channels and to describe the 
film as an interconnected network of atomic conducting paths \cite{ohring}. 

Bradley et al. \cite{bradley} were the first to propose and apply to the study
of EM a kinetic version of the random fuse model \cite{arcangelis}. The 
dynamic fuse model introduced by these authors \cite{bradley} adopts a failure
criterion for the elementary resistor of the network suitable for the 
description of the EM process. The predictions of this model concerning the 
damage pattern and TTFs (maximum and minimum TTF, relationship of the median 
time to failure with the current, the temperature and the line geometry) shows
the effectiveness and the potentiality of the resistor network approach 
\cite{bradley}. However, the model of Bradley et al. cannot describe many 
other features of EM. In fact this model, though giving a good description of 
the driving force of the electronic current, does not take into account the 
antagonist action exerted by mechanical stress gradients. On the other hand, 
the competition between these two effects is essential for giving rise to a 
threshold current for EM \cite{ohring,em_review}. Therefore, all the 
phenomenology related to the Blech's law  
\cite{ohring,em_review,filippi_96_02,blech} and saturation effects 
\cite{filippi_96_02} cannot be accounted for within the Bradley's model. 
Moreover, this model completely neglects Joule heating effects which are 
present in high stress condition \cite{ohring,em_review}.

Here, we illustrate a theoretical approach to EM which aims at studying the
different features associated with this phenomenon within a unified 
theoretical framework. Similarly to the approach of Bradley et al., 
our study is performed by renouncing to provide a description of the 
kinetics at an atomistic level and by adopting the RN approach, thus  
focusing on the correlations established by the electronic current among the 
different components of the system (grains, clusters of grains, interfaces, 
etc. i.e. atomic transport channels). However, in contrast to the 
dynamic fuse model \cite{bradley}, we use the biased percolation model 
\cite{gingl96,pen_prl_fail}, which adopts a probabilistic failure 
criterion for the elementary resistor of the network. The EM damage is 
described in terms of competition between two biased stochastic processes 
taking place in a resistor network \cite{pen_physd}. 
Then, by means of Monte Carlo (MC) simulations, we are able to study a variety
of relevant features of EM degradation. Early stage results has been presented
in Refs. [\onlinecite{pen_physd,pen_mcs}]. In this article, we present a 
comprehensive study which includes many fundamental new features.

As a first we will show results concerning the damage pattern, the resistance 
evolution and alloying effects. Then, large attention will be devoted to the 
behavior of TTFs. We will show that the model correctly predicts a lognormal 
distribution for them, perfectly superimposing with the experimental one. 
The dependence of the parameters of the TTF distribution on temperature and 
current has also been investigated. The Black's law 
\cite{ohring,em_review,black67} has been recovered not only for low but also 
for high current values and the current exponents are in good agreement with 
the experimental results in both cases. By using a rectangular geometry, the 
dependence of TTFs on the length and width of the metallic line has been 
investigated, so that, for a fixed width, the Blech's law 
\cite{ohring,em_review,blech} has been tested and a general expression has 
been obtained  for the dependence of TTFs  on both, length and width. Finally, 
but not of minor interest, we have considered resistance saturation effects 
\cite{filippi_96_02} and the properties of resistance fluctuations, by 
focusing on the non-Gaussianity of the distribution and on their power 
spectrum. Thus, our approach is able to account for a phenomenological
scenario much wider than that considered by all existing EM models.

Finally, we emphasize the fact that the interest in EM phenomena is not 
limited to strictly applicative and practical purposes. Indeed, the 
understanding  of non-equilibrium and failure phenomena in disordered
systems represents a fundamental topic which has attracted a large attention 
in the recent literature \cite{chakrabarti,stauffer,sahim98,torquato,arcangelis,gingl96,pen_prl_fail,lavine,eberhart,bardhan,politi02,sornette97,stan_zap,hansen,sornette92}. 
In this respect, it must be noted that EM, which occurs in granular materials, 
in presence of a significant disorder, driven by an external bias and 
contrasted by growth of mechanical stress gradients, exhibits practically
all the main ingredients to represents a paradigmatic example of failure 
process in a disordered system. 

The paper is organized as follows. Sec. II briefly surveys the theoretical 
model and define the parameters of interest. Sec. III presents the results 
in connection with: (A) resistance evolution, (B) stress conditions, 
(C) geometrical effects, (D) resistance saturation and fluctuations.  
Sec. IV draws the main conclusions.

\section{Model}
We describe a thin metallic line of length $L$, width $W$ and thickness $t_h$
as a two-dimensional resistor network of rectangular shape and 
square-lattice structure. The network of resistance $R$ is made
by $N_{L}$ and $N_{W}$ resistors in the length and width directions 
respectively. The external bias, represented by a constant voltage $V$ or 
a constant current $I$, is applied to the RN through electrical contacts 
realized by perfectly conducting bars at the left and right hand sides of the 
network. Thus, the total number of network resistors (excluding the contacts) 
is: $N_{tot}= 2N_{L}N_{W} + N_{L} -N_{W}$. Each resistor can be associated 
with a single grain, a small cluster of grains or an interfacial path. 
By denoting with $d$ the average size of the grains, of the grain clusters or 
of the interfacial path, the values of $N_{L}$ and $N_{W}$ can be related to 
the ratios $L/d$ and $W/d$, respectively. The network lies on an insulating 
substrate at temperature $T_0$, acting as a thermal bath, and it is made  by 
three kinds of resistors: i) regular resistors, ii) impurity resistors, 
iii) broken resistors. The regular resistors are associated with grains of 
``normal'' resistivity (void free). The resistance of these resistors depends 
linearly on temperature, according to the expression: 
\begin{equation} 
r_{reg,n} (T_n)=r_{ref}[ 1 + \alpha (T_n - T_{ref})]
\label{eq:temp_dep}
\end{equation}
Where $n$ is the resistor label, $\alpha$ the temperature coefficient of the
resistance (TCR), $T_n$ the local temperature, $T_{ref}$ and $r_{ref}$ the 
reference values for the TCR. When the Joule heating is negligible $T_n = T_0$ 
and the regular resistors are all equal to $r_0 \equiv r_{reg}(T_0)$.
The impurity resistors of resistance $r_{imp} < r_0$ are associated with the 
formation/dissolution of CuAl$_2$ precipitates (low-resistivity cluster) 
during the stress conditions. Thus, they account for the variation in the 
alloy composition during the EM test due to the electron wind and/or to 
thermal effects (alloying effects) \cite{scorzoni96,pen_physd,dekker,kao}. 
Finally, the broken resistors correspond to the presence of microvoids at the 
grain boundary and, possibly, inside the grains. These broken resistors of 
resistance $r_{OP}$ (OP stays for open circuit) are thus associated with very 
high resistivity regions inside the line. Here, we have taken 
$r_{OP} = 10^9 r_0$. 
The existence of temperature gradients due to current crowding and 
Joule heating effects is accounted for by taking the local temperature of the 
{\em n}-th resistor (regular, impurity or broken) of resistance $r_n$ given
by the following expression \cite{pen_prl_fail}:
\begin{equation}
T_n=T_{0} + A \Bigl[ r_n i_n^{2} + {B \over N_{neig}} \sum_{m=1}^{N_{neig}}
 \Bigl( r_{m,n} i_{m,n}^2   - r_n i_n^2 \Bigr) \Bigr],  \label{eq:temp}
\end{equation}
where $A$ is the thermal resistance of the single resistor, $N_{neig}$ is
the number of first neighbors of the {\em n}-th resistor, $i_n$ the current 
flowing in it and $i_{m,n}$ is the current flowing in the {\em m}-th neighbor. 
The value $B=3/4$ is chosen to provide uniform heating of the perfect network, 
i.e. made by identical resistors. Equation~(\ref{eq:temp}) is the Fourier 
equation written by taking the simplifying assumption of instantaneous
thermalization of the resistor, i.e. by taking a stationary regime and 
neglecting time dependent effects in the heat diffusion \cite{thermal_fuse}. 
Under this hypothesis, the diffusive term must balance the term related to the
power supplied by the external current and generated on the resistor or 
transmitted to it by mutual interactions.  A simplified form of 
Eq.~(\ref{eq:temp}), not including thermal exchanges with neighbor resistors 
has been firstly proposed by Gingl et al. \cite{gingl96}.
For a perfect \cite{note_perfect} or nearly perfect network of resistance  
$R_0$, when a mean-field approach is meaningful, the average temperature 
increase is thus \cite{pen_pre_fnl}:
\begin{equation}
\Delta T = A R_0 I^2 /N_{tot} \equiv \theta R_0 I^2,  \label{eq:thermalres}
\end{equation}
where $\theta = A /N_{tot}$ is the structure thermal resistance 
\cite{ohring,em_review}.

The EM damage, consisting in the formation of microvoids under the action of 
the electron wind, is simulated by a stochastic process of resistor breaking. 
In other terms, we consider that the transformation of the {\em n}-th resistor 
(regular or impurity) into a broken one, $r_{n} \rightarrow  r_{OP}$, can 
occur with probability $W_{OP,n}$. By adopting the biased percolation model 
\cite{gingl96,pen_prl_fail}, we have taken the following expression for 
$W_{OP,n}$: 
\begin{equation}
W_{OP,n}=\exp[-E_{OP}/k_B T_n],  \label{eq:w_break}
\end{equation}
where $E_{OP}$ is a characteristic activation energy. In fact, 
Equation ~(\ref{eq:w_break}) coupled with Eq.~(\ref{eq:temp}), 
implies that the void formation process is a biased percolation 
\cite{gingl96,pen_prl_fail}. This means a probability of breaking 
a resistor (generation of microvoids) higher for resistors crossed by high 
current values \cite{gingl96,pen_prl_fail}. Thus, the breaking probability is 
nonuniform for the different resistors in the network and it changes with 
time. We note that $W_{OP,n}$ depends on the current distribution, which in 
turn depends on the network configuration \cite{gingl96,pen_prl_fail}. As the 
last one results from a progressive accumulation of damage, the history of the
network is partly accounted for by the biased percolation model. 
 
The effect of the atomic back-flow, which contrasts the EM, is simulated by 
introducing a recovery process consisting in a stochastic healing of the 
broken resistors \cite{pen_physd,pen_pre_fnl,pen_upon99}. Therefore, the 
transformation $r_{OP} \rightarrow r_{reg,n}$ is allowed with probability: 
\begin{equation}
W_{R,n}=\exp[-E_{R}/k_B T_n],  \label{eq:w_recov}
\end{equation}
where $E_{R}$ is a recovery activation energy.
Furthermore, the variation of the line composition due to alloying effects
is described by allowing the stochastic transitions \cite{pen_physd}: 
$r_{reg,n}  \rightarrow r_{imp}$ and $r_{imp}  \rightarrow r_{reg,n}$,
occurring with probabilities $W_{RI,n}=\exp[-E_{RI}/k_B T_n]$ and 
$W_{IR,n}=\exp[-E_{IR}/k_B T_n]$ respectively, where $E_{RI}$ and $E_{IR}$
are two characteristic activation energies. By adopting this description
of alloying effects, we are limiting ourself to consider the change in the 
elementary resistances due to a variation inside any grain or cluster of 
grains (single resistor) of the number of Cu atoms dispersed in the matrix or 
present in small CuAl$_2$ precipitate \cite{pen_physd}. 

The initial configuration of the RN can contain some initial concentrations 
of broken and impurity resistors, $p_{ini}$ and $p^{imp}_{ini}$, 
respectively. Alternatively this initial configuration can be chosen as the 
perfect one: $p_{ini}=0$ and $p^{imp}_{ini}=0$ . 
The network evolution is obtained by Monte Carlo simulations which are
carried out according to the following iterative procedure.

(i) Starting from the initial network, we calculate $\{i_n\}$ and the network
resistance $R$ by solving Kirchhoff's loop equations. Moreover, we calculate 
$\{T_n\}$ by using Eq.~(\ref{eq:temp}). 

(ii) OP and $r_{imp}$ are generated with the corresponding probabilities 
$W_{OP}$ and $W_{RI}$ and the remaining $r_{reg}$ are changed according 
to $\{T_n\}$. Then $\{i_n\}$ and $\{T_n\}$ are recalculated. 

(iii) OP and $r_{imp}$ are recovered with probabilities $W_{R}$ and $W_{IR}$ 
respectively, and the resistances $r_{reg}$ are changed again according to 
$\{T_n\}$. 

(iv) $\{i_n\}$, $\{T_n\}$ and $R$ are recalculated. 

This procedure is iterated from (ii), thus the loop (ii)-(iv) corresponds to 
an iteration step, which is associated with a unit time step on an arbitrary 
time scale to be calibrated by comparison with experiments. 
Depending on the parameter values, which are related to the physical 
properties of the line and to the external conditions, the two following 
possibilities can be achieved during the iteration procedure: irreversible 
failure or steady-state evolution. In the first case, at least one 
percolating cluster of broken resistor (i.e. a cluster connecting the top and 
the bottom of the network) is formed and thus the resistance $R$ diverges 
\cite{stauffer}. In the second case, the network resistance fluctuates around 
an average value $<R>$ (saturation value) \cite{pen_pre_fnl,pen_upon99}. The 
average over the statistical ensemble (different realizations of failure of 
networks with the same parameters and in the same external conditions) of the 
values of the fraction of broken resistors corresponding to the appearance of 
at least one percolating cluster, is called percolation threshold and it is 
denoted as $p_c$  \cite{stauffer}.

To check the model we have considered EM tests performed with a standard 
median time to failure technique \cite{ohring,em_review} on Al-0.5\%Cu lines. 
The tests have been carried out at different currents and temperatures by 
adopting a 2-metal level configuration with tungsten vias 
\cite{ohring,em_review} and by using a 20\% relative resistance variation as 
failure criterion. The lines used in the tests were $3000$ $\mu$m long, $0.45$
$\mu$m wide and $0.8$ $\mu$m thick. The last thermal treatment undergone by 
these lines occurred during fabrication and it consisted in a high temperature
annealing followed by a rapid cooling. This treatment left a non-equilibrium 
concentration of Cu dissolved into the Al matrix. Therefore, in the early 
stage of the EM test, the heating associated with the stress conditions gives 
rise to a formation of CuAl$_2$ precipitates. The low resistivity of these 
clusters and, mainly, the reduction of the internal disorder, i.e. the 
reduction of the number of scattering centers of Cu in the solid solution, 
cause an initial decrease of the line resistance \cite{scorzoni96}. Here, 
we consider the data obtained at $T=492$ K, shown in 
Ref. \onlinecite{pen_physd}, and the data obtained at $T=467$ K, reported 
in the following section. In both cases the stress current density was 
$j=3$ MA/cm$^2$. The resistance of the lines at the reference 
temperature $T_{ref}=273$ K was: $R_{ref}^{line}=269$ $\Omega$ 
(averaged over a family of 40 samples), while the TCR was 
$3.6 \times 10^{-3}$ K$^{-1}$. 

The values of the parameters used in the simulations have been chosen as 
follows. We have taken the values corresponding to the actual stress conditions
and to the physical parameters of the metallic line whenever possible, i.e. 
whenever it was possible to make a direct correspondence with the model 
parameters and the line properties and when this choice was not too heavy
computationally. The remaining parameters have been chosen to fit the 
experimental results and/or to reduce the computational effort. 
Concerning this point, we notice that the present approach allows for a direct 
simulation of lines with ratio $L/W$ up to $\approx 150$. To describe the 
resistance evolution of lines characterized by higher values of this ratio, 
as the lines tested in the experiments shown in the next section 
(where $L/W=6667$), we have adopted  the following further approximation. 
The network is taken to represent the region of dominant void growth inside a 
longer line, i.e. the region responsible for the resistance variation of the 
line. We can take the length of this region given by $L/F$, where $F$ is an 
integer number. Thus, in the initial conditions: $R_{0,line} =  F R_{0}$. 
On the other hand, according to the above assumption we have: 
$\Delta R_{line} = \Delta R$. Therefore, the relative resistance variation 
of the whole line can be expressed as: 
$\Delta R_{line}/R_{0,line} = (1/F)(\Delta R/R_{0})$.
We underline that this approximation is used only to check the model by  
a direct comparison of the measured resistance evolutions of long lines 
(Fig. 2) and the  evolutions calculated by the present model (Fig. 3). 
All the other results concern short lines and do not make use of the above 
approximation.

Thus, except when differently specified, we have used the following values of 
the parameters: $N_W=12$,  $N_L=400$, $F=1$, $T_0=492$ K, 
$I=JWt_h=10.8$ mA (which corresponds to the values of $j$, $W$ and $t_{th}$
used in the EM tests cited above), $\alpha=3.6 \times 10^{-3}$ K$^{-1}$, 
$T_{ref}=273$ K, $r_{ref} = 0.048$ $\Omega$, $r_{imp}=0.016$ $\Omega$.
For the long lines used in the EM tests, when $F=200$, this value of $r_{ref}$
provides the correct value for $R_{ref}^{line}$, reported above. 
Moreover, we have taken $A=2.7 \times 10^8$ K/W. According to 
Eq.~(\ref{eq:thermalres}), this value of $A$ provides an initial heating of 
the network of $8.3$ K, comparable with that estimated in the experiments of 
Ref. \onlinecite{pen_physd}. The initial network configuration corresponds 
to $p_{ini}=0$ and $p^{imp}_{ini}=0$. Furthermore, we have taken: 
$E_{OP}=0.41$ eV and $E_{R}=0.35$ eV, as reasonable values for the 
activation energies $E_{OP}$ and $E_{R}$. We notice that the value of 
$E_{OP}$ controls the range of the time scale, nevertheless this range is in 
any case arbitrary within our model. Thus, the value of $E_{OP}$ can be 
considered as a free parameter which can be chosen with the purpose of saving 
computational time. More crucial is the choice of $E_{R}$, whose value sets 
the importance of the recovery process \cite{pen_pre_fnl}, i.e. the strength 
of the atomic back-flow due to the mechanical stress. Consequently with this 
choice of $E_{OP}$ and $E_{R}$, the values of $E_{RI}$ and $E_{IR}$ have been 
taken sufficiently small to account for the separation of the temporal scales 
of the void formation and of the alloying processes, observed in the 
experiments. Here, we have taken: $E_{RI}= 0.22$ eV and $E_{IR}=0.17$ eV. 

\section{Results} 
\subsection{Resistance evolution}
A typical resistance evolution and the corresponding damage pattern near the 
final failure are reported in Figs. 1(a) and 1(b), respectively. In this case 
$N_L = 48$ while all other parameters take the values specified at the end of 
the previous section. In Fig. 1(a) we observe a resistance drop at the early 
stages of the evolution due to the generation of impurity resistors. This 
process $r_{reg,n}  \rightarrow r_{imp}$ simulates the variation of 
composition of the line associated with the initial precipitation of part of 
the Cu dissolved in the Al matrix, as discussed in Sec. II. On the other hand,
this process together with the antagonist one, 
$r_{imp}  \rightarrow r_{reg,n}$, takes place on a time scale that is 
much shorter than that associated with the generation of broken resistors. 
Therefore, after a given amount of time (relaxation time of the alloying 
process), the concentration of the impurity resistors reaches its steady state
value  \cite{pen_physd} corresponding to the temperature $T_0 + \Delta T$. 
On a longer time scale, the fraction of broken resistors increases and the 
network becomes more and more unstable. This implies an increase of both the 
resistance value and the resistance fluctuations, as shown in Fig. 1(a). 
Finally, at a given time (time to failure) the fraction of broken resistors 
reaches the percolation threshold and $R(t)$ diverges. Figure 1(b) reports the
damage pattern just before the final failure. Precisely, this figure shows the
temperature distribution inside the network: the broken resistors are the 
missing ones while the different gray levels, from black (cold) to white 
(hot), correspond to different $T_n$  values ranging from the substrate 
temperature up to $700$ K, with a temperature  step of $10$ K. The damage 
pattern mainly consists of a channel of broken resistors elongated in the 
direction perpendicular to the current flow, a characteristic feature of the 
biased  percolation \cite{gingl96,pen_prl_fail}. This simulated damage pattern
reproduces well the experimental pattern observed by scanning electron or 
x-ray microscopy in metallic lines which are failed due to EM 
\cite{ohring,em_review,pattern_1,pattern_2}.

Figure 2 shows the resistance evolutions of seven Al-0.5\%Cu lines measured 
in the EM tests performed at $T=467$ K, as described in Sec. II. These lines 
were stressed by a current density $j=3$ MA/cm$^2$ which corresponds
to $I=10.8$ mA. Figure 3 reports the resistance evolutions obtained by 
simulations. Here, different curves correspond to different realizations of 
failure. In this case, we have taken: 
$N_W=12$, $N_L=400$, $F=200$, $T_0=467$ K, $r_{ref} = 0.044$ $\Omega$, 
$r_{imp}=0.006$ $\Omega$, $p_{ini}=(2.5 \pm 0.2) \times 10^{-2}$ (the broken 
resistors in the initial network configuration are supposed uniformly 
distributed), while the remaining parameters have the same values specified at 
the end of Sec. II. The time scale in Fig. 3 has been calibrated according to 
the following procedure. The statistical sample tested in the experiments was 
composed by thirteen lines and the resulting median time to failure, $t_{50}$,
was: $t_{50,exp} \approx 1.3 \times 10^6$ s. A sample of thirteen 
simulated failure realizations was considered and the corresponding 
$t_{50,sim}$ was calculated  in units of iteration steps. From these two
values, we obtained the value $\Delta t = 185$ sec for the time interval to be
associated with each iterative step. The comparison between Figs. 2 and 3
shows that the calculated evolution of the resistance well reproduces the main 
features of the observed evolution. We note that a small discrepancy between
measured and simulated evolutions appears just before the failure, where the 
abrupt increase of the resistance shown by the simulated curves, contrasts with
a pre-failure increase of the resistance generally present in the 
experimental ones. This discrepancy, is partly due to the factor $1/F$ 
relating the relative resistance variation of the network with that of the 
metallic line. In fact, the simulated evolutions obtained for short lines, 
see Fig. 1(a), when the full metallic line can be directly simulated and the 
approximation adopted for long lines can be avoided (i.e. $F=1$) display also 
a pre-failure region.  

The agreement is further confirmed by the comparison between the distributions
of the measured and calculated TTFs reported in Fig. 4. This figure shows on a 
lognormal plot \cite{lognor_plot} the cumulative distribution function (CDF) 
of the  failure probability as a function of the times to failure obtained 
from the EM tests (full circles) and simulations (open circles) considered 
above. The agreement between experiments and simulations is excellent and the 
shape factor of the two distributions is $s=0.16$ in both cases. Thus, the 
direct comparison of the results of the model and of the EM tests shown in 
Figs. 2, 3 and 4, validate the present computational approach for the study 
of EM failures.  

As anticipated in Sec. I, two central problems encountered in the study of 
EM phenomenon concern the role played by the stress conditions and the
line geometry on the damage process \cite{ohring,em_review,blech}. Therefore, 
with the purpose of further checking the predictivity of the model and 
of extracting new information from it, we have calculated the effect
on TTFs of temperature, stress current, length and width of the lines. 
To contain the computational effort and to avoid the approximation used for 
long lines, the study has been limited to short or moderately short lines. 
Thus, in the following we will discuss the results of simulations carried out 
by taking $F=1$, by varying $T_0$, $I$, $N_L$, $N_W$, while keeping all the 
remaining parameters to the values specified at the end of Sec. II. 

\subsection{Stress conditions: temperature and current effect}
We start by considering the effect of temperature on the times to failure
of $20$ networks of sizes $12 \times 400$ stressed by a current of $I=10.8$ 
mA. Accordingly, we analyse the dependence on temperature of $t_{50}$
and $s$, the two parameters which determine a lognormal distribution. We
consider fourteen values of $T_0$ ranging from 400 K to 800 K, a considerably
wider range with respect to standard accelerated tests 
\cite{ohring,em_review}. Figure 5 shows the cumulative distribution functions 
of the failure probability calculated for $T_0=800$ K (triangles left), 
$T_0=650$ K (open squares), $T_0=467$ K (open circles) and $T_0=400$ K 
(triangles down), while the solid lines fit the CDFs with lognormal 
distributions. These four $T_0$ values are selected as representative for the 
behavior of $s$. Indeed, we have found that $s$ is nearly independent of 
temperature in a wide range of intermediate temperature values, while it 
increases significantly at low temperatures and decreases at the highest 
temperatures considered here. The broadening of the distribution 
at low temperatures and its narrowing at high temperatures witness the 
different importance of the network microgeometry in the two extreme stress 
conditions of $400$ and $800$ K, respectively. Indeed, the network 
microgeometry, resulting from the stochasticity of the defectiveness, give 
rise to a kind of network individuality. At very high stress, the differences 
in the network microgeometry loose their importance. The contrary occurs at 
low stress, where this diversity becomes of importance in determining the 
actual TTF. We underline that the broadening of the TTF distribution at low 
temperatures has important implications in the interpretation and use of the 
results of accelerated EM tests. In fact, when evaluating the reliability of a
family of lines under standard operating conditions (usually close to room 
temperature and relatively low current density), it is crucial to estimate not
only $t_{50}$ but also the minimum time to failure 
\cite{filippi_jap78,filippi_apl66}. The determination of these two quantities 
is usually obtained from accelerated tests performed at high temperatures on 
a statistically significant, but in any case small, sample of the entire 
family. Then, the estimate of the minimum time to failure of the
family is obtained by an extrapolation of the CDF in the region of low failure 
probability \cite{filippi_jap78,filippi_apl66}. Such an estimate is very 
sensitive to a possible broadening of the TTFs distribution at the operation 
temperature. For this reason it is crucial to estimate and to take into 
account the dependence of $s$ on temperature. We remark that the increase of 
$s$ at low temperatures is a source of the following apparent paradox 
\cite{filippi_apl66}: the minimum time to failure at low temperatures can be 
shorter than the minimum time to failure at high temperatures. We will face a 
similar paradox by discussing the dependence of $s$ on the current 
\cite{filippi_apl66}. Solutions to this apparent paradox has been proposed in 
the literature \cite{filippi_apl66,capasso}, based on the necessity of 
testing large samples and on the introduction of a three parameters lognormal 
distribution, where the third parameter is a characteristic incubation time 
\cite{filippi_apl66}.

The analysis of the temperature dependence of the TTFs is completed by Fig. 6 
which displays on linear-log scale the calculated values of $t_{50}$ as a 
function of the inverse of the substrate temperature. Here, the dashed line is 
the best fit obtained with the function $Z \ exp[E/k_B T_0]$, where $Z$ is
a fitting amplitude. Thus, the calculated values of $t_{50}$ perfectly follow,
within the numerical uncertainty, the Black's law 
\cite{ohring,em_review,black67}, discussed in Sec. I. The value of $E$ 
extracted from the fit is $E=0.41$ eV, thus $E = E_{OP}$, and we can identify 
the activation energy of the resistor breaking process with the EM activation 
energy.

To investigate the effect of current on the failure process, we have 
calculated the TTFs of $12 \times 400$  networks stressed at $T_0=492$ K by 
different (thirteen) current values in the range $5.0 \div 60.$ mA. For each 
current value $s$ and $t_{50}$ have been determined by considering twenty
realizations of failure. Figure 7 shows the CDFs of the failure probability 
versus failure time, calculated for $I=7.5$ mA (triangles down), $I=30.$ mA
(open squares), $I=60.$ mA (open circles). The solid lines fit the CDFs with 
lognormal distribution. We have found that the shape factor of the distribution
exhibits a minimum at intermediate values of $I$. Thus, we can identify two 
regions of current values: a moderate current (m.c.) region, where $s$ 
decreases at increasing current, and a high current (h.c.) region, where $s$ 
increases at increasing current. Such a broadening of the TTFs distribution at 
low currents has been actually observed in several EM tests 
\cite{filippi_jap78,filippi_apl66}. Its implications concerning the evaluation
of reliability of metallic lines are similar to those previously mentioned in 
connection with the effect of temperature. A detailed discussion of these 
problems can be found in Ref. \onlinecite{filippi_apl66}. Here, we underline 
that the non-monotonic behavior of $s$ versus current contrasts the general 
monotonic behavior found versus temperature. This non-monotonic behavior of 
$s$ can be understood by considering also the dependence of $t_{50}$ on the 
current that is reported in Fig. 8. More precisely, we show in the inset of 
this figure a log-log plot of the calculated values of $t_{50}$ versus $I$. 
We can see that $t_{50}$ exhibits two power-law regions. A first one is in 
the moderate current region and is characterized by a current exponent 
$n=2.1$, in good agreement with the Black's law \cite{shatzkes86,black67}. 
A second power-law with a higher exponent, $n=5.7$, is found in the high 
current region. We notice that similar behaviors at high current densities 
have actually been observed in EM measurements and are frequently reported in 
the literature \cite{ohring,em_review,nat_IC}. Furthermore, the same inset of 
Fig. 8 shows that $t_{50}$ drastically increases at the lowest currents. 
Here, for $I=4.0$ mA the networks are found to remain stable over
more than $5 \times 10^5$ iterations. This strong increase of $t_{50}$ is 
associated with the existence of a threshold current, $I_B$, below which a
steady state condition is achieved, manifesting itself in a saturation of the 
network resistance. Accordingly, for $I<I_B$ the electrical breakdown no longer
occurs. For this reason, in Fig. 8 the region corresponding to $I<I_B$ is
evidenced in gray. Previous investigations of the general properties of the 
model, reported in Refs. \onlinecite{pen_pre_fnl}, have shown the existence
of this threshold. Some of the properties of the steady state of the
network will be discussed later in connection with the results concerning 
resistance noise and we refer the reader to Refs. \onlinecite{pen_pre_fnl} 
for a deeper analysis of these properties. We emphasize that this sharp 
increase of $t_{50}$ at low currents has also been observed in EM tests 
\cite{em_review,filippi_jap78}. Therefore, the dependence of $t_{50}$ on the 
current obtained by simulations agrees with the behavior measured over the 
full range of current values. To complete the analysis, Fig. 8 reports in a 
log-log plot the median time to failure versus the difference $I-I_B$ (full 
squares). By taking for $I_B$ the value given above, $I_B=4.0$ mA, we have 
found that $t_{50}$ scales as:
\begin{equation} 
t_{50} \sim (I-I_B)^{-n_g}
\label{eq:black_gen}
\end{equation}
This expression, firstly proposed by Filippi et al. \cite{filippi_jap78}, can 
be considered as a generalization of the Black's law and $n_g$ can be called
as generalized current exponent, to distinguish it from the current exponent 
$n$ of the conventional Black's law, Eq.~(\ref{eq:Black}). 
Figure 8 displays two current regions, each characterized by a given value of 
$n_g$. These two regions correspond to the different current dependence of the 
distribution shape factor, reported in Fig. 7. The common behavior of $t_{50}$ 
and $s$ with current originates from the change with the bias of both the
damage pattern and the magnitude of Joule heating and can be understood
as follows.
 
First, let us consider a simpler system: a network in which two stochastic 
processes of resistor breaking and recovery occur with uniform probabilities:  
$W_{D0}=\exp[-E_{OP}/k_B T_0]$ and $W_{R0}=\exp[-E_{R}/k_BT_0]$. A similar
network, subjected to random percolation \cite{stauffer,pen_prl_stat} describes
well the instability of very thin metallic films due to agglomeration 
phenomena \cite{alford}. The stability of this network  has been 
studied in Ref. \onlinecite{pen_prl_stat}, where the failure condition and 
the expression for the average time to failure \cite{def_attf} (ATTF) have 
been derived in the limit of networks of infinite size. Here, to point out 
the dependence of the percolation threshold on the system size, it is 
convenient to write the failure condition in the following form:
\begin{equation}
W_{D0} > {p_c \over(1-p_c)}{W_{R0} \over(1 - W_{R0})}
\label{eq:fail_cond}
\end{equation}
similarly, the average time to failure can be written as:    
\begin{equation} 
ATTF={\ln(1-q)  \over \ln[(1-W_{D0})(1-W_{R0})] }
\label{eq:attf}
\end{equation}
with:
\begin{equation} 
q \equiv p_c \Bigl [1 + {W_{R0} \over W_{D0}(1-W_{R0})} \Bigr ]
\label{eq:qdef}
\end{equation}
where $q<1$ for failing networks, according to Eq. ~(\ref{eq:fail_cond}). For 
this simple system it is quite easy to determine the role of the size and of 
the geometry of the network on the value of the percolation threshold. In 
fact, an ideally infinite network ($N_L \rightarrow \infty$ and
$N_W \rightarrow \infty$), with percolation threshold $p_{c,\infty}$, 
would have a zero probability of breaking for a fraction of defects $
p<p_{c,\infty}$ and a breaking probability equal to 1 for $p>p_{c,\infty}$ 
\cite{stauffer}. However, for networks of finite size there is a non vanishing
probability of formation of the percolating cluster of defects (thus, of 
breaking) also when the defect fraction is $p<p_{c,\infty}$ and, 
for contrast, a probability less than 1 for $p>p_{c,\infty}$ \cite{stauffer}.
For this reason, when networks of finite size are considered, $p_c$ is defined
as the average over the statistical ensemble of the minimum values of $p$ 
corresponding to the formation of at least one percolating cluster 
\cite{stauffer}, as anticipated in Sec. II. In the case of $N \times N$ 
networks with $N$ finite, it has been found \cite{stauffer} that 
$(p_c - p_{c,\infty}) = c N^{-1/\nu}$, where $\nu$ is the correlation length 
exponent (with universal value $\nu=4/3$ in two-dimensions), while the 
proportionality constant, $c$, depends on the lattice structure. 
In particular, for networks with square-lattice structure, it has been found 
\cite{ziff} that $c \approx 0$ and thus $p_c \approx p_{c,\infty}=0.5$, 
independently of $N$. In the case of rectangular $N_W \times N_L$ networks 
with square-lattice structure, we have found that for a fixed value of $N_W$, 
$p_c$ decreases with $N_L$, by reaching its minimum value, $p_{W,\infty}$, 
when $N_L \rightarrow \infty$. More precisely: 
$(p_c - p_{W,\infty}) \sim N_L^{-1/2\nu}$, where  $p_{W,\infty}$ decreases
with $N_W$. The reduction of $p_c$ when increasing the length, by keeping 
constant the width of the network, can be explained as a consequence of the 
existence of a higher number of possible paths of defects connecting the top 
and the bottom of such a network compared to a square network with the same 
width. These feature is associated with the greatest instability of networks 
with this geometry.
 
We note that Eqs.~(\ref{eq:fail_cond})-(\ref{eq:qdef}) and the subsequent 
discussion apply to the case of random percolation, therefore the average 
time to failure given by Eq.~(\ref{eq:attf}) is independent of the current. 
On the other hand, electromigration is a current driven phenomenon which, 
within our approach, is described by a biased percolation 
\cite{gingl96,pen_prl_fail}. The effect of the biased percolation can be 
roughly decomposed in two components : i) a correlated growth of the defect 
pattern, which exhibits an increasing degree of filamentation at increasing 
currents; ii) an average heating of the network due to Joule effect. 

The filamentation driven by the external current implies a bias dependent 
percolation threshold $p_c(I)$, as shown in Fig. 9. In this figure the full 
circles represent the values of the percolation threshold calculated by 
averaging the fraction of defects which corresponds to the failure of 20 
networks of sizes $12 \times 400$ subjected to external currents $I>I_B$. 
The gray region in Fig. 9 corresponds to values  $I<I_B$ (stationary region), 
while the solid curve is the best-fit with a quadratic expression. We notice 
that for low currents the defect pattern is found to exhibit a relatively
week degree of filamentation and the value of $p_c=0.21$ is not very far
from the random percolation threshold (which, for these values of $N_W$ and 
$N_L$, is $p_c=0.37$). Then, for currents up to $I \approx 40$ mA the 
filamentantion achieves its maximum level and consequently $p_c$ its minimum 
value. At further increasing $I$ values, the onset of a multi-filamentation 
pattern is observed \cite{pen_varna} with the simultaneous growth of several 
filaments of defects (voids) elongated perpendicular to the direction of the 
current flow \cite{pen_varna}. This effect manifests itself in a smooth 
increase of the percolation threshold at the highest currents, as shown 
in Fig. 9. 

By introducing the above dependence of  $p_c$ on $I$ in 
Eqs. ~(\ref{eq:attf})-~(\ref{eq:qdef}) we obtain the average time to failure 
versus current shown in Fig. 10 (up-triangles). For the sake of comparison, 
we also show in this figure (open circles) the values of $t_{50}$ obtained by 
MC simulations and already reported in the inset of Fig. 8. We can see that 
at low currents the ATTFs obtained from Eq. ~(\ref{eq:attf}) by accounting for
the bias driven filamentation, tend to merge with the MC results. 
Then, in the moderate current region the ATTF exhibits a power-law behavior, 
with a current exponent $n=1.5$ (the dashed curve in Fig. 10 is the best-fit 
with such a power-law). Finally, for currents $I \approx 40$ mA the ATTF 
nearly saturates. The discrepancy at intermediate and high currents between 
the two sets of data (ATTFs from Eq. ~(\ref{eq:attf}) and MC simulations) can 
be explained in terms of Joule heating effects. These effects can be included 
in Eq. ~(\ref{eq:attf}), in the spirit of a mean field theory, by replacing 
in the random percolation expressions of the breaking and recovery 
probabilities, the temperature $T_0$ with an average temperature 
$<T>=T_0 + \Delta T$, where $\Delta T$ is given by Eq. ~(\ref{eq:thermalres}).
By introducing into Eq. ~(\ref{eq:attf}) these new average probabilities 
$<W_{OP}>$ and $<W_R>$ and by simultaneously accounting for the dependence 
$p_c(I)$, we obtain the average time to failure reported as down-triangles in 
Fig. 10. Thus, the up triangles in this figure correspond to a value of the 
structure thermal resistance $\theta=0$ and the down triangles to  
$\theta \ne 0$. The excellent agreement found between MC results and those 
obtained through Eq. ~(\ref{eq:attf}) supports the interpretation suggested 
here. Thus, the generalization of Eq. ~(\ref{eq:attf}) to the case of biased 
percolation, made by accounting for both Joule heating effects and for the 
dependence of the percolation threshold on the bias, is able to describe quite
well both the results of the MC simulations and the behavior observed in many 
EM experiments \cite{ohring,em_review,nat_IC,filippi_jap78}.

The results reported in  Fig. 10 shed new lights on the role played by  
Joule heating effects in determining the value of the current exponent 
in the Black's law. Indeed, though many accelerated EM tests 
\cite{ohring,em_review,nat_IC} provides a value $n=2$ \, other values of $n$ 
have been frequently measured \cite{ohring,em_review,nat_IC,tammaro99}. 
Values of $n$ greater than $2$ have been usually attributed to Joule heating 
\cite{ohring,em_review,nat_IC,tammaro99}. In this respect, Fig. 10 points out
quite well the importance of this effect in the high current region, thus 
confirming the interpretation made in the literature. However, Fig. 10 shows 
that Joule heating can affect the value of $n$ also in the moderate current 
region. In this region values $1<n<2$ have been reported by different authors 
\cite{ohring,em_review,tammaro99}. Many EM models have been proposed to 
explain the value of the current exponent 
\cite{ohring,em_review,shatzkes86,tammaro99}. These models fall into two 
main categories: ``void growth'' and ``nucleation'' models \cite{tammaro99}. 
In the void growth models, the failure is taken to occur after a void grows 
up to a critical size. It is generally accepted that this category of models 
provides a value $n=1$ \cite{tammaro99}. In nucleation models, the failure 
arises from the buildup of a critical vacancy concentration. It is generally 
believed that this second category of models provides a value $n=2$ 
\cite{shatzkes86}. Recently, Tammaro and Setlik \cite{tammaro99} have proved 
that even when nucleation is the limiting process it can be $1<n<2$. Our 
results confirm this conclusion. As a matter of fact that the present model, 
which relates the failure to the achievement of a percolation threshold for 
the defect fraction, belong to the second category of models. In fact, as 
shown by Fig. 10, metallic lines with different structure thermal resistance 
$\theta$, thus exploiting different sensitivity to Joule heating effects can 
exhibit different current exponent values, even in the moderate current 
region. Concerning this point, we underline the fact that a power-law 
behavior of the ATTF versus current is predicted by the model merely as a 
consequence of the filamentation of the damage pattern, even neglecting the 
effect of the average heating of the metallic line. Moreover, the dependence 
of $n$ on the length of the lines, reported in EM tests \cite{nat_IC}, can be
explained in terms of dependence of both, heating effects and $p_c$, on the 
length of the system. Finally, from Fig. 10 we can see that Joule heating is 
also responsible for the shift towards lower current values of the crossover 
between the high-current and the moderate-current regions (in this case from 
$40$ mA to $30$ mA). On the basis of the above considerations we can now 
understand the dependence with current of the shape factor of the TTFs 
distribution shown in Fig. 7. At relatively low currents, just above the 
$I_B$ threshold, the degree of filamentation of the defect pattern is week 
and there is a wide spread in the network microgeometries and thus in the 
TTFs. Then, at increasing currents, the degree of filamentation increases and
the effect of the different network microgeometries is reduced together with 
the spread in the TTFs. Finally, in the highest current region, because of 
strong Joule heating, multi-filamentation occurs \cite{pen_varna}. This 
implies a high degree of stochasticity and in turn a high variability of 
microgeometries together with an increase of the TTFs spreading.

\subsection{Geometrical effects}
In this section we investigate the effect of the length and width of the 
network on the failure process. First, we have analysed the dependence 
of $t_{50}$ on the length $N_L$ of the network. Figure 11 displays the 
simulated median time to failure as a function of $N_L$. The data indicated by
full circles are obtained by taking: $N_W = 12$ and $I=10.8$ mA, while the 
data reported as open squares and shown in the inset of Fig. 11 correspond to 
$N_W = 36$ and $I=32.4$ mA. Thus, in both cases the current density (measured 
in current units) \cite{cur_dens} is $j=I/N_W=0.9$ mA. We have found that 
$t_{50}$ sharply increases by decreasing $N_L$ and it diverges for network 
lengths below a certain value, $N_{L_c}$. This length can be considered as a 
critical length of the network and it is dependent on the network width, as 
shown in Fig. 11. Furthermore, for sufficiently long networks, $t_{50}$ nearly
saturates to a value independent of the length and increasing when increasing 
the width. In the following, this asymptotic value of the median time to 
failure in the limit of infinitely long lines will be denoted as $t_{inf}$.
The behavior of $t_{50}$ shown in Fig. 11 is in qualitative good agreement 
with the behavior observed in the EM experiments \cite{em_review}.  

To determine more precisely the dependence of the median time to failure on 
the geometrical parameters, we have analysed the simulated values of $t_{50}$
as a function of the difference $(N_L - N_{L_c})$. Figure 12 reports the 
log-log plot of the difference $t_{50}-t_{inf}$ as a function of the 
difference $(N_L - N_{L_c)}$. Here, together with the data already shown 
in Fig. 11 we have also reported the data for $N_W=48$ and $I=43.2$ mA 
(open circles). In this way, all the data in Fig. 12 correspond to the same 
value of $j$. We have found the following values of critical length: 
$L_c=2.5, 7.5, 10.0$, for $N_W=12, 36, 48$, respectively. Therefore, in all 
the cases it is: $(N_{L_c}/N_W) = 0.21 \pm 0.01$. The values of $t_{inf}$
range between $2 \times 10^3 \div 3 \times 10^3$. Figure 12 shows that the 
difference $t_{50}-t_{inf}$ follows a power-law as a function of the
difference $(N_L - N_{L_c})$; indeed, the solid, dashed and long dashed 
curves in this figure fit the data with a power-law with exponent $-\lambda$, 
where $\lambda=0.62 \pm 0.02$. In particular, as shown in the inset of 
Fig. 12, we have found that the three sets of data collapse onto the 
same curve once the difference $t_{50}-t_{inf}$ is considered as a 
function of the normalized quantity $(N_L - N_{L_c})/ N_{L_c}$. Thus,
the difference $t_{50}-t_{inf}$ scales with the ratio 
$(N_L - N_{L_c})/ N_{L_c}$ :
\begin{equation} 
t_{50} -t_{inf} \sim  \Bigl[ {N_L - N_{L_c} \over N_{L_c}} \Bigr]^{-\lambda} 
\sim \Bigl[ {N_L \over N_{L_c}} - 1 \Bigr]^{-\lambda}
\label{eq:t50_scal}
\end{equation}
With the aim of identifying the physical parameters determining the value of  
$\lambda$, we have investigated the dependence of $t_{50}$ on the length of 
the networks when these are stressed by different current densities. The
results of simulations are reported in Figs. 13 (a) and (b). Figure 13(a) 
displays the difference $t_{50}-t_{inf}$ versus $(N_L - N_{L_c})$ for 
$N_W=36$ and $j=0.9$ mA (open squares) and $j=0.75$ mA (up-triangles). 
The first set of data is the same of Fig. 12 (thus $N_{L_{c1}}=7.5$), while 
for the second set it is $N_{L_{c2}}=8.9$ and therefore $N_{L_c}/N_W = 0.25$. 
First we note that it is: 
\begin{equation} 
(jN_{L_c})_1 \approx (jN_{L_c})_2
\label{eq:Blech_network}
\end{equation}
in agreement with the Blech's law \cite{ohring,em_review,blech} (in this case 
we have obtained for the threshold product a value $6.8$ mA ). Second, we have 
found that the value of $\lambda$ is different for the two sets of data. 
Precisely, for the data obtained by taking $j=0.75$ mA it is 
$\lambda=1.04 \pm 0.03$. 

Figure 13(b) displays the difference $t_{50}-t_{inf}$ versus $(N_L - N_{Lc})$ 
for $N'_W=48$. The current densities $j=0.9$ mA (open circles) and $j=0.75$ mA 
(down-triangles) are the same of those in Fig. 13(a), while the values of the 
critical lengths are respectively: $N'_{L_{c3}}=10.0$ and $N'_{L_{c4}}=12.0$. 
The value of the exponent found for $j=0.75$ mA is now $\lambda=1.20 \pm 0.04$.
This result suggests a dependence of the exponent $\lambda$ not only on $j$.
We note that also for this set of data the Blech's condition, 
Eq.~(\ref{eq:Blech_network}), is satisfied with ${(jN_L)_c}'=9.0$ mA. Thus, 
by considering the threshold product as a function of the different 
parameters, ${(jN_L)_c}=F(r_0,N_W,E_R,T_0...)$, $F$ is found to be an 
increasing function of $N_W$.

According to the model proposed by Blech \cite{blech}, the threshold product 
is determined by the ratio $(\Omega_a \Delta \sigma)/(\rho Z^*e)$ 
(Eq.~(\ref{eq:Blech}) in Sec. I). The quantity $(\rho Z^*e)$ is related to the
intrinsic properties of the material and, within our model, can be associated 
with the parameter $r_0$, defined in Sec. II. On the other hand, the product 
$(\Omega_a \Delta \sigma)$, being a function of the geometry of the line, 
of the properties of the electrical contacts, of the presence of passivation
layers and of the temperature \cite{ohring,em_review,filippi_96_02,blech}, 
within our model is controlled by the geometry of the network, by the 
efficiency of the recovery process (i.e. the energy $E_R$) and by the 
temperature $T_0$. In particular, according to Eq.~(\ref{eq:Blech}), the value
of the threshold product $(jL)_c$ is an increasing function of $\Delta \sigma$
which is the maximum value of the mechanical stress difference between the 
line terminals that the line can bear \cite{ohring,em_review,blech}. 
On the other hand, the value of this maximum stress difference increases with 
the line width. Therefore, the dependence of the threshold product on the 
width predicted by our model is in qualitative agreement with the behavior 
described by Eq.~(\ref{eq:Blech}). In any case, further studies are necessary 
to determine the dependence of $t_{inf}$, $\lambda$ and of the threshold 
product on $r_0$, $N_W$, $E_R$, $T_0$, and the other parameters of the model.

Below we analyse in more detail the dependence of the simulated median time to
failure on the width of the network. To this purpose Fig. 14 reports $t_{50}$ 
versus $N_W$ for networks of different lengths, stressed by a current density 
$j=0.9$ mA. Going from the top to the bottom of the figure the different 
sets of data correspond to: $N_L=15, 18, 36, 48, 100$, respectively. The 
different curves in Fig. 14 fit the corresponding data with the expression:  
\begin{equation} 
t_{50} = K \Bigl[ \Bigl({N_L - N_{L_c} \over N_{L_c}} \Bigr)^{-\lambda}  
- \tau \Bigr] F\Bigl(N_W/N_{W0}\Bigr)
\label{eq:fit_wdep}
\end{equation}
where $F(x) \equiv (1 - e^{-x})$, $\lambda=0.62$, $N_{L_c} = 0.21 N_W$, 
$W_0 = 7 \div 12$, $\tau$ is a constant related to the value of $t_{inf}$, 
and $K$ is a fitting constant. As a general trend, $t_{50}$ increases 
systematically at increasing the width of the network. However, in the case of
long lines the dependence of $t_{50}$ on the width shows a saturation at the 
largest width. By contrast, in the case of short lines this saturation 
disappears and $t_{50}$ exhibits a final sharp increase with the width and a 
tendency to diverge when approaching the Blech condition. This behavior of 
$t_{50}$ is in overall agreement with the results of EM tests performed on 
lines larger than about $2$ $\mu$m \cite{ohring,em_review}. 
On the other hand, it has been found that narrow lines, with width smaller 
than $2$ $\mu$m, exhibit a sharp lengthening of the median time to failure 
\cite{ohring,em_review,nat_IC,cho}. This phenomenon occurs because for such 
narrow lines (known as bamboo structures) the grain size becomes comparable 
with the line width \cite{ohring,em_review}. As a consequence, the lack in 
these lines of grain boundaries along the direction of current implies that 
EM can only occur within the bulk of the grains or along interfaces. 
These processes usually require an activation energy significantly higher 
than that associated with EM along grain boundaries \cite{ohring,em_review}.
Therefore, bamboo structures exhibit median time to failure longer than that 
displayed by other kinds of lines \cite{ohring,em_review,nat_IC,cho}. 
Figure 14 shows that our model, in the present formulation, is unable to 
describe the lengthening of $t_{50}$ for very narrow lines. This is not 
surprising considering the fact that here we have taken a single value for the
activation energy of the breaking process, $E_{OP}$, that is a value common to
all the resistors in the network. A further implementation of the model which 
introduces two different activation energies for the breaking of bulk 
resistors and surface resistors, $E_{OP,B}$ and $E_{OP,S}$ respectively, with 
$E_{OP,B} > E_{OP,S}$, would account also for this lengthening of $t_{50}$ for
narrow lines typical of bamboo structures. 

\subsection{Saturation and resistance fluctuations}
In this section we consider the situation occurring at low current densities
or for short line lengths, when the product of the current density and of the 
line length is lower than the threshold value. In this case electromigration
stops and the resistance of the line achieves a steady state value (saturation
value) dependent on the external conditions and on the properties of the 
line \cite{blech,filippi_96_02}.
The steady state is characterized by fluctuations of the resistance around the
average value $<R>$ (this value is calculated by averaging over all the steady
state values of the resistance, i.e. the values taken after the transient time
associated with the termination of the EM process) \cite{pen_pre_fnl}. The 
study of resistance saturation effects and of their dependence on the stress 
current, temperature, geometry and other properties of the metallic line, can 
provide an important tool to investigate EM phenomena \cite{filippi_96_02}, 
alternative to the study of the median time to failure. Actually, studies of 
EM based on saturation effects exhibit the advantage to be non-destructive and
in particular, they reveal their effectiveness in the analysis of short 
lines \cite{filippi_96_02}. Furthermore, from a fundamental point of view, 
they provide the possibility to investigate fluctuation phenomena under far 
from equilibrium conditions.   

Below we consider the steady state of networks biased by currents below the
electrical breakdown threshold $I \le I_B$. Figure 15 shows the resistance 
evolution of a $12 \times 400$ network with the same parameters considered 
in Sec. III.B. More precisely, we report the difference 
$R(t) - R_0$ versus time for a network stressed at $T_0=492$ K by the current 
$I=4.0$ mA which corresponds to the threshold for electrical failure. 
Saturation to $<R>$ and large resistance fluctuations are well evident in 
Fig. 15 for this current value. The inset in this figure reports on an 
enlarged time scale the initial transient values of $R -R_0$, evidencing also 
the initial decrease of the resistance due to alloying effects, discussed in 
Sec. II. Details concerning the dependence of $<R>$ on the value of the 
stress current, on the bias conditions (constant current or constant voltage) 
and on the TCR can be found in Refs. \onlinecite{pen_pre_fnl}. 

Here we discuss two important features of the resistance fluctuations 
occurring in a metallic line in a non-equilibrium steady state, stressed by
a current near the EM threshold. First, we consider the 
non-Gaussianity property \cite{weissman,bloom,bramwell_prl} of the 
distribution of $\delta R \equiv R \ - <R>$. To this purpose, we have 
calculated the probability density function, $\Phi$, of the distribution 
of $\delta R$ for the steady state signal in Fig. 15. Figure 16 reports 
the product $\Phi \sigma$ as a function of $(<R>-R)/\sigma$ 
in a lin-log plot, where $\sigma$ is the root mean square deviation from the 
average resistance. For comparison, in the same figure we also report the 
Gaussian distribution (dashed curve), which in this normalized representation 
has zero mean and unit variance. This representation has been adopted because,
by making the distribution independent of its first and second moments, it is 
particularly convenient for exploring the functional form of a 
distribution \cite{bramwell_prl}. The results in Fig. 16 show a considerable 
non-Gaussianity of the distribution of $\delta R$, which is associated with 
the fact that the system is close to breakdown conditions. We have found 
that this non-Gaussianity is sufficiently well fitted by a generalized form 
\cite{bramwell_prl} of the Gumbel distribution (solid curve), where this last 
is commonly used for the analysis of extreme events \cite{ohring}. The role 
of the stress current and of the size of the system on the distribution of the
resistance fluctuations has been studied in Refs. \cite{pen_ngauss}, where the
conditions under which the distribution achieves the universal behavior 
described by the Bramwell, Holdsworth and Pinton distribution 
\cite{bramwell_prl} have been identified.

As a second feature, we consider the power spectral density of 
resistance fluctuations. To this purpose, Fig. 17 displays the power spectrum 
associated with the steady state signal in Fig. 15. Both the frequency and 
the spectral density are expressed in arbitrary units. Two regions can be 
identified in the spectrum: a 1/f like branch at low frequencies and a 
Lorentzian cut-off at high frequencies. This feature is here interpreted as 
due to the presence of three relaxation mechanisms: i) the slowest one related
to the breaking and recovery of the regular resistors and describing the 
generation and recovery of microvoids; ii) the fastest one associated with the
transformation of regular resistors into impurity resistors (and vice versa) 
and describing alloying effects; iii) an intermediate one corresponding to 
generation and breaking of impurity resistors. Accordingly, the Lorentzian 
branch describes the evolution occurring on a fast scale when only the second 
transformation is present. On the slow scale, the other two transformations 
become of importance and thus all the relaxation mechanisms coexist. 
Therefore, a 1/f-like spectrum appears in the low frequency region, in 
agreement with resistance noise measurements performed on Al-Cu alloys 
\cite{em_review,koch,jones}. We notice that by taking different activation
energies for bulk and surface resistors, as described at the end of 
Sect. III.C, a further relaxation time would be added. 

\section*{Conclusions}
We have applied the biased percolation model to the study 
of degradation and failure phenomena induced by EM in metallic line. Our 
``coarse grain'' approach focuses on the correlations established by the 
electronic current among the different elemental resistors of the network
which mimic the structural components of the system (grains, clusters of 
grains, interfaces, etc. i.e. atomic transport channels). This approach 
provides a unified theoretical framework able to account successfully for 
many relevant features of the experiments, including the damage pattern, 
the resistance evolution, alloying effects and the statistical properties 
of TTFs. In particular, the model correctly predicts a lognormal 
distribution for TTFs, perfectly  superimposing with the experimental one
and it is able to estimate the dependence of the shape factor of the
distribution on current and temperature. In what concerns the dependence of 
the median time to failure on the stress conditions, the model predictions 
agree with the experiments over the full range of current and temperature 
values considered. Simulations performed on rectangular networks of different 
length and width have allowed us to investigate the dependence of TTFs on 
these parameters. The results of the model agree with the existence of a 
Blech's length, moreover they predict the existence of a scaling relation 
between the MTF and the line length. Finally, we have considered resistance 
saturation effects. In this case we have studied the properties of the 
resistance fluctuations, by focusing on the non-Gaussianity of the 
distribution and on their power spectrum. The flexibility of the theoretical 
approach offers further possibility to describe and interpret phenomena which 
at present have not been considered. We finally emphasize that EM, which 
occurs in granular materials, in presence of a significant disorder, driven 
by an external bias and contrasted by growth of mechanical stress gradient, 
represents a paradigmatic example of failure process in a disordered system. 
Therefore, the ability of our approach to account for a wide scenario of 
the EM related phenomenology should be of interest in the more extensive 
perspective of understanding non-equilibrium and failure phenomena in 
disordered systems. 

\section*{ACKNOWLEDGMENTS}
This work has been performed within the STATE project of the INFM. 
Support from the cofin-03 project ``Modelli e misure di rumore in 
nanostrutture'' financed by Italian MIUR and from SPOT-NOSED 
project IST-2001-38899 of EC is also gratefully acknowledged.

\vfill\break\noindent

FIGURE CAPTIONS

\vspace{0.5truecm}\noindent
Fig. 1 (a).  Typical resistance evolution of a $12 \times 48$ network. 
The values of all the other parameters are specified at the end of Sec. II.
The stress conditions are: $I=10.8$ mA and $T_0=492$ K. The resistance is 
expressed in Ohm and the time in arbitrary units corresponding to the number of
iteration steps. (b) Damage pattern at the iteration step $t=2150$ of the 
evolution shown in (a). The different gray levels, from black to white, are 
associated with different $T_n$ values, ranging from $492$ to $700$ K with 
a step of $10$ K.

\vspace{0.5truecm}\noindent
Fig. 2. Experimental resistance evolutions of seven Al-0.5\%0.5Cu lines 
stressed at $T=467$ K by a current density $j=3$ MA/cm$^2$ (which 
corresponds to $I=10.8$ mA within the model). The resistance is
expressed in Ohm and the time in seconds.

\vspace{0.5truecm}\noindent
Fig. 3. Calculated resistance evolutions of the Al-0.5\%0.5Cu lines in Fig. 2.
The simulations have been performed by taking $T_0=467$ K and $I=10.8$ mA.
The value of the remaining parameters are specified in the text. The 
resistance is expressed in Ohm and the time in seconds by using the value  
$\Delta t = 185$ s for the time interval associated with each 
iterative step (see text).

\vspace{0.5truecm}\noindent
Fig. 4. Lognormal plot of the cumulative distribution function of the
failure probability (expressed in percentage) as a function of the time to
failure: (full circles) TTFs experimentally measured and (open circles) 
calculated by the model. The data correspond to the same statistical samples 
considered in Fig. 2 and 3, respectively. The stress conditions are $T_0=467$ 
K and $I=10.8$ mA. The dashed line fits the CDFs with a lognormal 
distribution. 

\vspace{0.5truecm}\noindent
Fig. 5. Lognormal plot of the cumulative distribution functions of the 
failure probability (expressed in percentage) as a function of the time to
failure. The different functions are calculated at different substrate
temperatures: $T_0=800$ K (triangles left), $T_0=650$ K (open squares), 
$T_0=467$ K (open circles), $T_0=400$ K (triangles down). The stress 
current is $I=10.8$ mA. The solid lines fit the CDFs with lognormal 
distributions.

\vspace{0.5truecm}\noindent
Fig. 6. Median time to failure, $t_{50}$, as a function of the inverse 
substrate temperature. The median times to failure are expressed in arbitrary 
units and the temperature in K. The stress current is $I=10.8$ mA. 
The dashed line is the fit with the exponential function $Z \ exp[4700/T_0]$. 

\vspace{0.5truecm}\noindent
Fig. 7. Lognormal plot of the cumulative distribution functions of the 
failure probability (expressed in percentage) as a function of the time to
failure. The different functions are calculated at different stress currents: 
$I=7.5$ mA (triangles down), $I=30.0$ mA (open squares), $I=60.0$ mA 
(open circles). The substrate temperature is $T_0=492$ K. The solid lines 
fit the CDFs with lognormal distributions.

\vspace{0.5truecm}\noindent
Fig. 8. Log-log plot of $t_{50}$ versus $I - I_b$, where $I_b$ is the 
breakdown current defined in the text. The two lines of slope $-1.5$ and 
$-5.2$ represent the fits with a power-law in the moderate current and in the 
high current regions, respectively. The inset shows the log-log plot of 
$t_{50}$ versus $I$. In both the main figure and the inset, the median times
to failure are expressed in arbitrary units and the current in mA. The gray 
region in the inset corresponds to the stationary region attainable for 
currents lower than the $I_B$ value. 
 
\vspace{0.5truecm}\noindent
Fig. 9. Percolation threshold for broken resistors, $p_c$, versus current 
(this last is expressed in mA). The curve is a quadratic fit (see text). 
The gray region evidences the stationary region.

\vspace{0.5truecm}\noindent
Fig. 10. Log-log plot of $t_{50}$ versus $I$. The median times to failure 
are expressed in arbitrary units and the current in mA. The open circles
(from MC simulations) represent the same data reported in the inset of 
Fig. 8. The up-triangles are obtained by Eq.~(\ref{eq:attf}) but taking
into account the dependence $p_c(I)$ shown in Fig. 9. The down-triangles
are obtained with the same procedure but replacing the probabilities  
$W_{D0}$ and $W_{R0}$ with $<W_{OP}>$ and $<W_R>$. The solid, long-dashed and 
dotted curves represent the best-fit with a power-law with slopes $-2.1$,
$-1.5$, $-5.7$, respectively.
  
\vspace{0.5truecm}\noindent
Fig. 11. Median time to failure, $t_{50}$, as a function of the network length
$N_L$ for $N_W=12$ and $N_W=36$ (inset). The median time to failures are 
expressed in arbitrary units. The dashed lines are a guide to the eyes.

\vspace{0.5truecm}\noindent
Fig. 12. Log-log plot of the difference $t_{50}-t_{inf}$ versus the 
difference $(N_L - N_{L_c)}$, where $t_{inf}$ is the median time to failure 
in the limit of infinitely long lines and $N_{L_c}$ is the critical length. 
Both quantities are expressed in arbitrary units. The full circles 
are obtained by taking $N_W=12$, $I=10.8$ mA. In this case is $N_{L_c}=2.5$. 
The open squares correspond to $N_W=36$, $I=32.4$ mA and $N_{L_c}=7.5$ while 
the open circles to $N_W=48$, $I=43.2$ mA and $N_{L_c}=10.0$. The solid, 
dashed and long dashed curves fit the data with a power-law of exponent 
$-0.62 \pm 0.02$. In the inset the same data of $t_{50}-t_{inf}$ are reported
as a function of $(N_L - N_{L_c})/ N_{L_c}$.

\vspace{0.5truecm}\noindent
Fig. 13 (a). Log-log plot of the difference $t_{50}-t_{inf}$ versus the 
difference $(N_L - N_{L_c)}$, where $t_{inf}$ is the median time to 
failure in the limit of infinitely long lines and $N_{L_c}$ is the critical 
length. Both quantities are expressed in arbitrary units. The network width is
$N_W=36$. The open squares are obtained by taking $I=32.4$ mA, in this case 
$N_{L_c}=7.5$. The up-triangles correspond to $I=27.0$ mA and $N_{L_c}=8.9$. 
(b) $N_W=48$; open circles:  $I=43.2$ mA and $N_{L_c}=10.0$; down-triangles: 
$I=36.0$ mA and $N_{L_c}=12.0$.

\vspace{0.5truecm}\noindent
Fig. 14. Plot of $t_{50}$ as a function of network width $N_W$. The 
different sets of data correspond to networks of different length: 
$N_L=15$ (down triangles), $N_L=18$ (open squares), $N_L=36$ (full squares), 
$N_L=48$ (up triangles), $N_L=100$ (stars). The curves fit the corresponding
data with the expression specified in the text. 

\vspace{0.5truecm}\noindent
Fig. 15. $R(t) - R_0$  versus time for a network $12 \times 400$ stressed at 
$T_0=300$ K by a current $I=I_B=4$ mA. Here $R_0$ is the perfect network 
resistance. The inset shows on an enlarged time scale the evolution of the 
resistance in the initial stage.

\vspace{0.5truecm}\noindent
Fig. 16. Normalized probability density function, $F\equiv \Phi \sigma$, of 
the resistance fluctuations reported in Fig. 16. The solid curve fits the 
data with a generalized Gumbel distribution (see the text), while the dashed 
curve is the Gaussian distribution.

\vspace{0.5truecm}\noindent
Fig. 17. Power spectral density of the resistance fluctuations in Fig. 16. 
The frequency and the spectral density are expressed in arbitrary units.


\begin{thebibliography}{10}
%
\bibitem{ohring} M. Ohring {\em Reliability and Failure of Electronic
Materials and Devices}, Academic Press, San Diego (1998). 

\bibitem{em_review}
A. Scorzoni, B. Neri, C. Caprile, and F. Fantini, {\em Mat. Science Rep.} 
{\bf 7}, 143 (1991) and F. Fantini, J. R. Lloyd, I. De Munari and 
A. Scorzoni, {\em Microelectronic Engineering} {\bf 40}, 207 (1998).

\bibitem{nat_IC} M. A. Alam, R. K. Smith, B. E. Weir, P.J. Silverman,  
{\em Nature} {\bf 420}, 378 (2002) and 
{\em Materials, Technology and Reliability for Advanced 
Interconnects and Low-k Dielectrics}, Ed. by K. Maex, Y. C. Joo, 
G. S. Oehrlein, S. Ogawa, J. T. Wetzel, Mat. Res. Soc. Symp. Proc., 
{\bf 612} (2000).


\bibitem{cho} J. Cho, C. V. Thompson, {\em Appl. Phys. Lett.} 
{\bf 54}, 2577 (1989).

\bibitem{healing}
Z. Li, C. L. Bauer, S. Mahajan and A. G. Milnes, {\em J. Appl. Phys.} 
{\bf 72}, 1821 (1992); S. Ohfuji and M. Tsukada, {\em J. Appl. Phys.} 
{\bf 78}, 3769 (1995); N. Stojadinovic, I. Manic, S. Djoric-Veljkovic, 
V. Davidovic, D, Dankovic, S. Golubovic, S. Dimitrijev,  
{\em Microelectron. Reliab.}, {\bf 42}, 1465 (1999).

\bibitem{koch} R. H. Koch, {\em Phys. Rev.} B, {\bf 48}, 12217 (1993) and
K. Dagge, W. Frank, A. Seeger, H. Stoll, {\em Appl. Phys. Lett.} 
{\bf 68}, 1198 (1996) and A. M. Yassine and C. T. M. Chen
{\em IEEE Trans. on Electronic Devices}, {\bf 44}, 180 (1997). 

\bibitem{filippi_jap78} R. G. Filippi, G. A. Biery and R. A. Wachnik,
{\em J. Appl. Phys.} {\bf 78}, 3756 (1995).

\bibitem{filippi_apl66} R. G. Filippi, G. A. Biery and R. A. Wachnik,
{\em Appl. Phys. Lett.} {\bf 66}, 1897 (1995).

\bibitem{filippi_96_02} R. G. Filippi, R. A. Wachnik, H. Aochi, J. R. Loyd, 
M. A. Korhonen, {\em Appl. Phys. Lett.} {\bf 69}, 2350 (1996) and 
R. G. Filippi, R. A. Wachnik, C. P. Eng, D. Chidambarrao, P. C. Wang, 
J. F. White, M. A. Korhonen, T. M. Shaw, R. Rosenberg, T. D. Sullivan, 
{\em J. Appl. Phys.} {\bf 91}, 5787 (2002).

\bibitem{scorz_j96}
A. Scorzoni, I. De Munari, H. Stulens and V.D'Haeger, J. Appl. Phys.,
{\bf 80}, 143, (1996).

\bibitem{scorzoni96}
A. Scorzoni, I. De Munari, R. Balboni, F. Tamarri, A. Garulli and 
F. Fantini, {\em Microelectron. Reliab.}, {\bf 36}, 1691 (1996).

\bibitem{foley98}
S. Foley, A. Scorzoni, R. Balboni, M. Impronta, I. De Munari, 
A Mathewson, and F. Fantini, {\em Microelectron. Reliab.}, 
{\bf 38}, 1021 (1998) and A. Scorzoni, S. Franceschini, R. Balboni, 
M. Impronta, I. De Munari, and F. Fantini, {\em Microelectron. Reliab.}, 
{\bf 37}, 1479 (1997).

\bibitem{jones} 
J. Guo, B. K. Jones, G. Tref\'an, {\em Microelectron. Reliab.} {\bf 39}, 
1677 (1999) and V. Dattilo, B. Neri, C. Ciofi, 
{\em Microelectron. Reliab.} {\bf 40}, 1323 (2000).

\bibitem{capasso} M. Gall, C. Capasso, D. Jawarani, R. Hernandez, H. Kawasaki
and P.S. Ho, {\em J. Appl. Phys.} {\bf 90}, 732 (2001) and 
{\em Appl. Phys. Lett.} {\bf 76}, 843 (2000.)

\bibitem{pattern_1} J. S. Huang, T. L. Shofner and J. Zhao, 
{\em J. Appl. Phys.}  {\bf 89}, 2130 (2001) and
G. Schneider, D. Hambach and B. Niemann, B. Kaulich, 
J. Susini, N. Hoffmann, W. Hasse, {\em Appl. Phys. Lett.}  {\bf 78}, 
1936 (2001).

\bibitem{pattern_2} B. C. Valek, J. C. Bravman, N. Tamura, A. A. MacDowell,
R. S. Celestre and H. A. Padmore, R. Spolenak, W. L. Brown, B. W. Batterman 
and J. R. Patel, {\em Appl. Phys. Lett.}  {\bf 81}, 4168 (2002).

\bibitem{shatzkes86}
M. Shatzkes and J. R. Lloyd {\em J. Appl. Phys.} {\bf 59}, 3980 (1986).

\bibitem{lloyd91}
J. R. Lloyd, and J. Kitchin, {\em J. Appl. Phys.} {\bf 69}, 2117, 1991.

\bibitem{korhonen93} M. A. Korhonen, P. Borgesen, D. D. Brown and C. Y. Li, 
{\em J. Appl. Phys.} {\bf 74}, 4995 (1993) and M. A. Korhonen, P. Borgesen, 
K. N. Tu and C. Y. Li, {\em J. Appl. Phys.} {\bf 73}, 3790 (1993).

\bibitem{bradley}
R. M. Bradley and K. Wu {\em Phys. Rev.} E {\bf 50}, R631 (1994);
K. Wu and R. M. Bradley, {\em Phys. Rev.} B {\bf 50}, 12468 (1994);
K. M. Crosby and R. M. Bradley, {\em Phys. Rev.} B {\bf 56}, 8743 (1997);
R. M. Bradley, M. Mahadevan and K. Wu, {\em Phyl. Mag.} B {\bf 79}, 257 (1999).

\bibitem{knowlton97} B. D. Knowlton, J. J. Clement and C. V. Thompson, 
{\em J. Appl. Phys.} {\bf 81}, 6073 (1997); 
J. J. Clement, {\em J. Appl. Phys.} {\bf 82}, 5991 (1997);
B. D. Knowlton and C. V. Thompson, {\em J. Mater. Res.} {\bf 13}, 1164 (1998).

\bibitem{krug98}
M. Schimschak and J. Krug, {\em Phys. Rev. Lett.} {\bf 80}, 1674 (1998) and
M. R. Gungor and D. Maroudas, {\em Appl. Phys. Lett.} {\bf 72}, 3452 (1998).

\bibitem{tammaro99} M. Tammaro and B. Setlik, {\em J. Appl. Phys.} {\bf 85}, 
7127 (1999) and
A. S. Oates, {\em Appl. Phys. Lett.} {\bf 66}, 1475 (1995).

\bibitem{proost} S. A. Chizhik, A. A. Matvienko, A. A. Sidelnikov, J. Proost, 
{\em J. Appl. Phys.} {\bf 88}, 3301 (2000).

\bibitem{sasagawa} K. Sasagawa, M. Hasegawa, M. Saka and H. Ab\'e,  
{\em J. Appl. Phys.} {\bf 91}, 1882 (2002).

\bibitem{black67}
J. R. Black, in {\em Proc. of 5th IEEE International Reliability Physics 
Symposium} p.148, (1967).

\bibitem{blech}
I. A. Blech, {\em J. Appl. Phys.} {\bf 47}, 1203 (1976);y
I. A. Blech and C. Herring, {\em Appl. Phys. Lett.} {\bf 29}, 132 (1976).

\bibitem{pen_physd}
C. Pennetta, L. Reggiani, G. Tref\'an, F. Fantini, A. Scorzoni
and I. De Munari {\em J. Phys. D: Appl. Phys.} {\bf 34}, 1421 (2001)
and C. Pennetta, L. Reggiani, G. Trefan, F. Fantini,
A. Scorzoni and I. De Munari, {\it Computational Material Science}, 
{\bf 22}, 13 (2001).

\bibitem{dekker} J. P. Dekker, C. A. Volkert, E. Arzt and P. Gumbsch, 
{\em Phys. Rev. Lett.} {\bf 87}, 035901 (2001).

\bibitem{kao} H. K. Kao, G. S. Cargill, C. K. Hu, {\em J. Appl. Phys.} 
{\bf 89}, 2588 (2001).

\bibitem{chakrabarti} B.K. Chakrabarti, L. Benguigui, 
{\em Statistical Physics of fracture and Breakdown in Disordered Systems}, 
Oxford Univ. Press, Oxford (1997); H. J. Herrmann and S. Roux, 
{\em Statistical Models for the fracture of disordered media}, North-Holland, 
Amsterdam (1990).

\bibitem{stauffer}
D. Stauffer and A. Aharony, Introduction to Percolation Theory,
(Taylor and Francis, 1992).

\bibitem{sahim98} M. Sahimi, {\em Phys. Rep.} {\bf 306}, 213 (1998).

\bibitem{torquato}
S. Torquato, {\em Random Heterogeneous Materials, Microscopic and Macroscopic 
Properties}, Springer-Verlag, New York, 2002.

\bibitem{arcangelis} L. De Arcangelis, S. Redner and H. J. Herrmann, 
{\em J. Phys. Lett.} (France) {\bf 46}, 585 (1985); L. De Arcangelis, 
A. Hansen, H. J. Herrmann and S. Roux, {\em Phys. Rev.} B {\bf 40}, 877 (1989).

\bibitem{gingl96}
Z. Gingl, C. Pennetta, L. B. Kiss, and L. Reggiani, 
{\em Semic. Sci. Technol.}, {\bf 11}, 1770 (1996).

\bibitem{pen_prl_fail}
C. Pennetta, L. Reggiani and G. Tref\'an, {\em Phys. Rev. Lett.}, {\bf 84}, 
5006 (2000) and C. Pennetta, L. Reggiani and L. B. Kish, 
{\em Physica} A {\bf 266}, 214 (1999).

\bibitem{pen_mcs}
C. Pennetta, L. Reggiani and  E. Alfinito, {\em Mathematics and
Comput. in Simulation}, {\bf 62}, 495 (2003).
 
\bibitem{lavine}
M. Lavine,  {\em Science}, {\bf 303}, 314 (2004).

\bibitem{eberhart}
M. E. Eberhart, {\em Why Things Break, Understanding the World by the Way It 
Comes Apart}, Harmony Books, New York, 2003.

\bibitem{bardhan} C.D. Mukherjee, K.K. Bardhan and M.B. Heaney,
{\em Phys. Rev. Lett.} {\bf 83}, 1215 (1999) and C. D. Mukherjee, 
K. K. Bardhan, {\em Phys. Rev. Lett.} {\bf 91}, 025702 (2003).

\bibitem{politi02} A. Politi, S. Ciliberto and R. Scorretti, 
 {\em Phys. Rev.} E {\bf 66}, 026107 (2002) 

\bibitem{sornette97} J. V. Andersen, D. Sornette and K.T. Leung,
 {\em Phys. Rev. Lett.} {\bf 78}, 2140 (1997) and 
L. Lamaign\`ere, F. Carmona and D. Sornette, {\em Phys. Rev. Lett.} 
{\bf 77}, 2738 (1996).

\bibitem{stan_zap} S. Zapperi, P. Ray, H. E. Stanley and 
A. Vespignani, {\em Phys. Rev. Lett.} {\bf 78}, 1408 (1997).

\bibitem{hansen} A. Hansen, S. Roux and E. L. Hinrichsen,
{\em Europhys. Lett.} {\bf 13}, 517, (1990); Y. Yagil, G. Deutscher and 
D. J. Bergman, {\em Phys. Rev. Lett.} {\bf 69}, 1423 (1992).

\bibitem{sornette92} 
D. Sornette and C. Vanneste, {\em Phys. Rev. Lett.}{\bf 68}, 612, (1992);
C. Vanneste and D. Sornette, J. Phys. I (France), {\bf 2}, 16212, (1992).

\bibitem{thermal_fuse} 
Time dependent effects on the heat diffusion have been studied by
D. Sornette and C. Vanneste in Ref. \onlinecite{sornette92}. In fact, 
Eq.~(\ref{eq:temp}) can be obtained from the expression used by 
Sornette et al. by assuming an instantaneous thermalization of each resistor 
and by adding the contribution of the power dissipated on first neighbor 
resistors.

\bibitem{note_perfect} The resistance $R_{perf}$ of a perfect network made by 
elementary resistors of resistance $r$ is given by the following expression: 
$R_{perf} = r {N_L \over N_W +1}$.

\bibitem{pen_pre_fnl}
C. Pennetta,  L. Reggiani, G. Tref\'an, E. Alfinito, {\em Phys. Rev.} E, 
{\bf 65}, 066119 (2002) 
and Pennetta C., {\em Fluctuation and Noise Letters}, {\bf 2}, R29 (2002).

\bibitem{pen_upon99}
C. Pennetta, G. Tref\'an, L. Reggiani, in {\em Unsolved Problems of Noise 
and Fluctuations}, Ed. by D. Abbott, L. B. Kish, AIP  Conf. Proc. 
{\bf 551}, New York (1999).

\bibitem{lognor_plot} In a lognormal plot, the stochastic variable (the
failure time in this case) is represented on a logarithmic scale while a
suitable transformation is applied to the coordinates of the axis
representing the cumulative distribution function in such a manner that a
lognormal distribution appears as a straight line. This kind of representation 
is thus particularly helpful to evaluate the lognormality of a distribution
and, for this reason, it is usually adopted in reliability analysis.
Further details can be found in Ref. \onlinecite{ohring}.

\bibitem{pen_prl_stat}
C. Pennetta, G. Tref\'an and L. Reggiani, {\em Phys. Rev. Lett.}, {\bf 85}, 
5238 (2000).

\bibitem{alford}
K. Sieradzki, K. Bailey, T. L. Alford, {\em Appl. Phys. Lett.} {\bf 79}, 
3401 (2001) and H. C. Kim, T. L. Alford, D. R. Allee, 
{\em Appl. Phys. Lett.} {\bf 81}, 4287 (2002).

\bibitem{def_attf} 
This quantity, defined as the arithmetical average of TTFs is different from 
$t_{50}$ although its value is close to that of $t_{50}$.

\bibitem{ziff}
R. M. Ziff, {\em Phys. Rev. Lett.} {\bf 69}, 2670 (1992).

\bibitem{pen_varna}
C. Pennetta, L. Reggiani and  G. Tref\'an, {\em Mathematics and
Comput. in Simulation}, {\bf 55}, 231 (2001).

\bibitem{cur_dens}
In our approach a thin line is described as a two-dimensional network, thus
the current density must be associated with quantity $I/N_W$. 

\bibitem{weissman} 
M. B. Weissman, {\em Rev. Mod. Phys.} {\bf 60}, 537, (1988).

\bibitem{bloom} N. Vandewalle, M. Ausloos, M. Houssa, P. W. Mertens, 
M. M. Heyns, {\em Appl. Phys. Lett.} {\bf 74}, 1579 (1999) and
I. Bloom, I. Balberg, {\em Appl. Phys. Lett.} {\bf 74}, 1427 (1999).
 
\bibitem{bramwell_prl} S. T. Bramwell, K. Christensen, J. Y. Fortin, P. C. W.
Holdsworth, H. J. Jensen, S. Lise, J. M. Lopez, M. Nicodemi, J. F. Pinton
and M Sellitto, {\em Phys. Rev. Lett.} {\bf 84}, 3744 (2000); 
S.T. Bramwell, P.C.W. Holdsworth, J. F. Pinton, {\em Nature}, {\bf 396}, 
552 (1998).

\bibitem{pen_ngauss}
C. Pennetta, E. Alfinito, L. Reggiani, S. Ruffo, 
{\em Semicond. Sci. Technol.}  {\bf 19}, S164 (2004); C. Pennetta, 
E. Alfinito, L. Reggiani, S. Ruffo, Cond-mat/0401352, {\em Physica A}, 
in print.

\end{thebibliography}
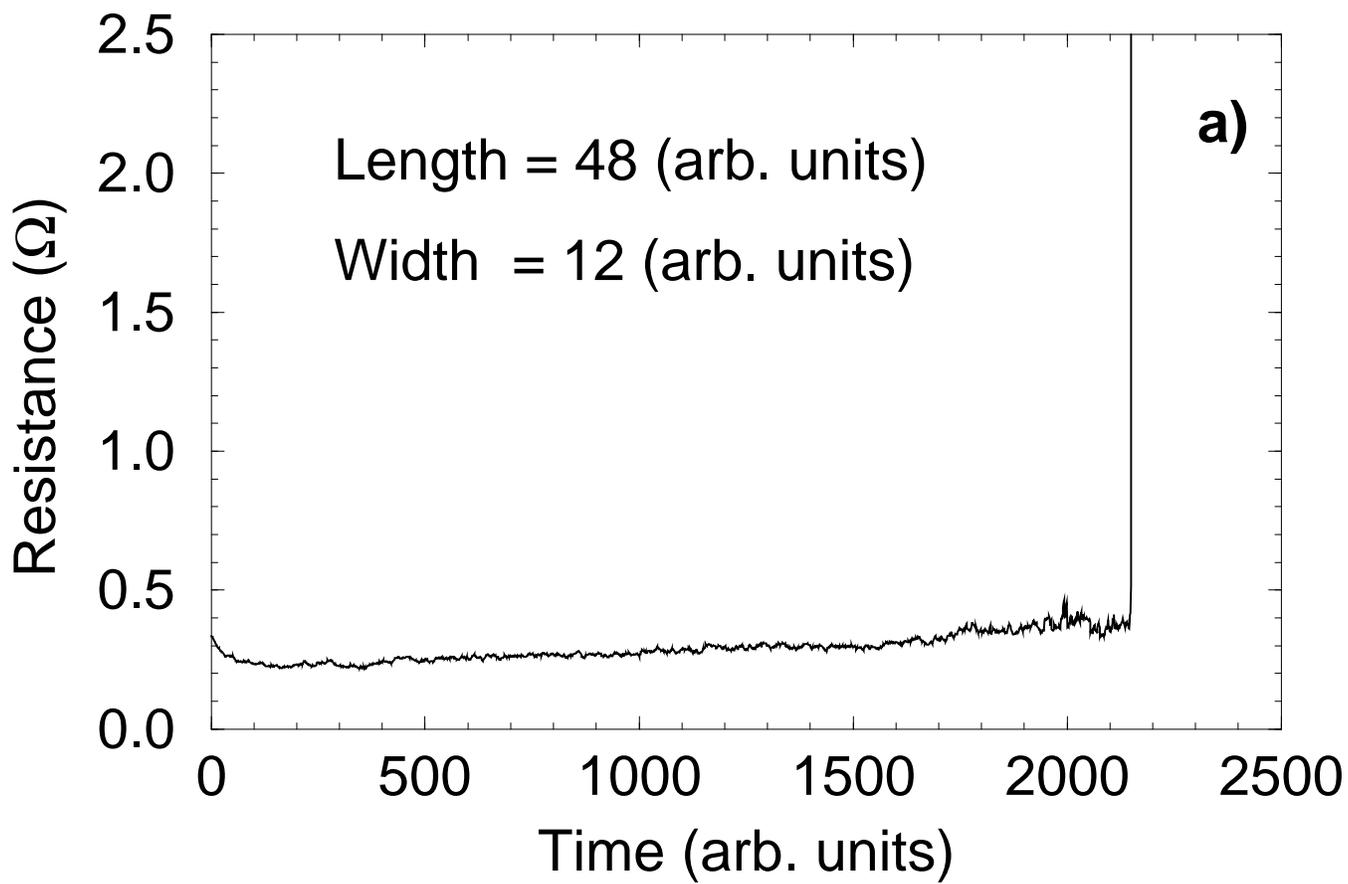
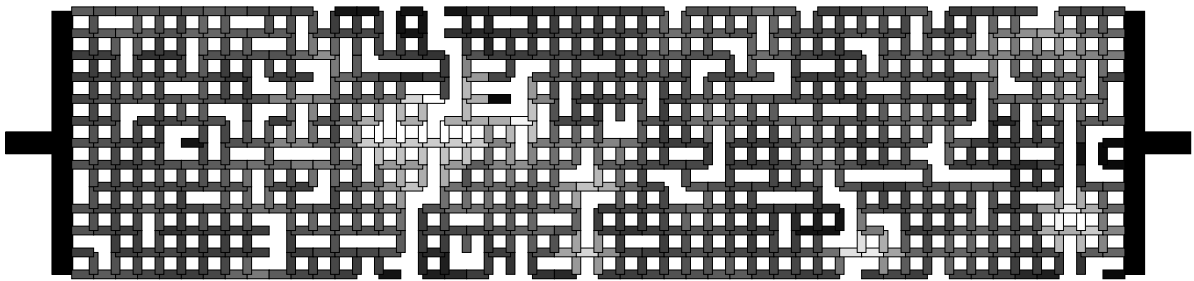
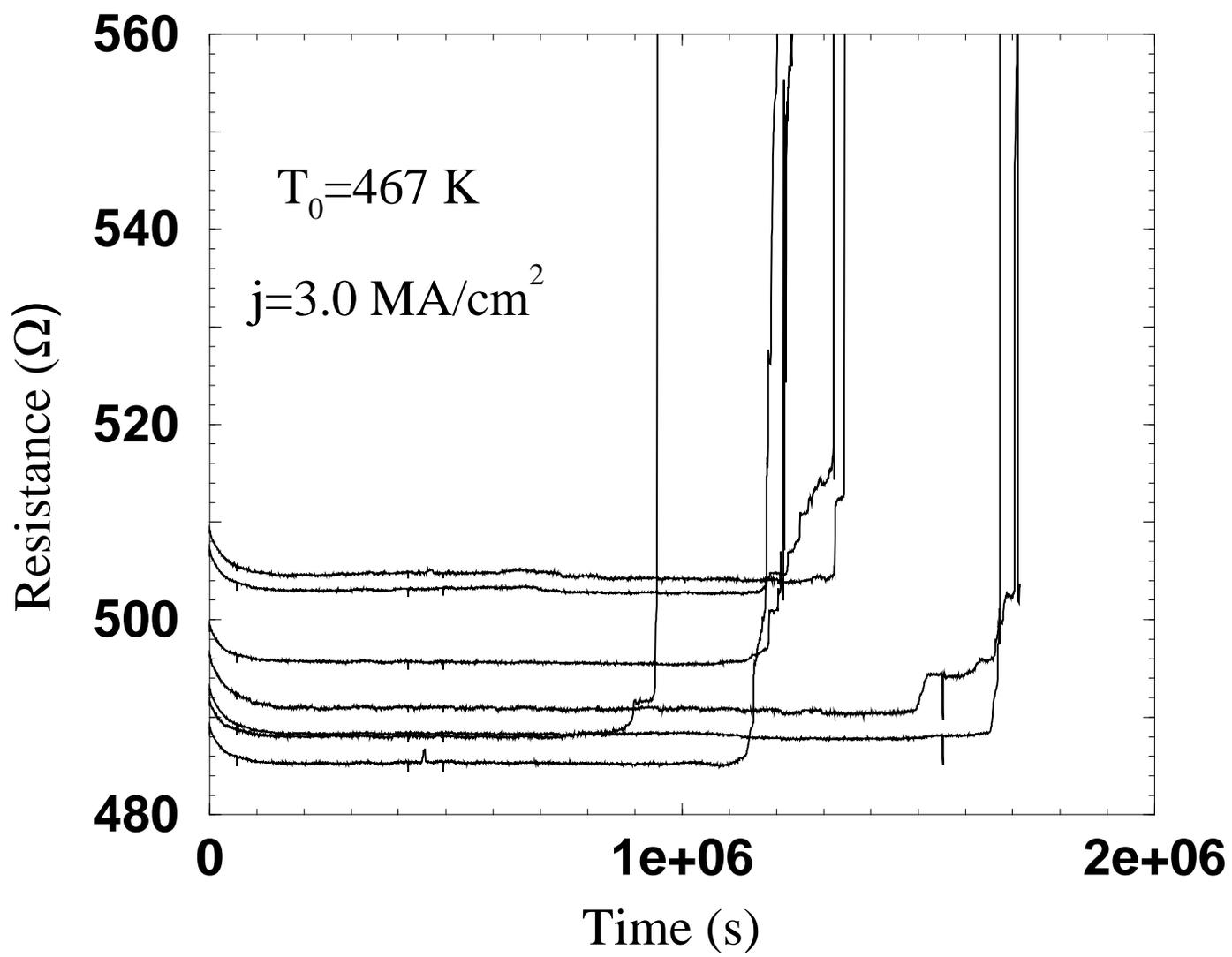
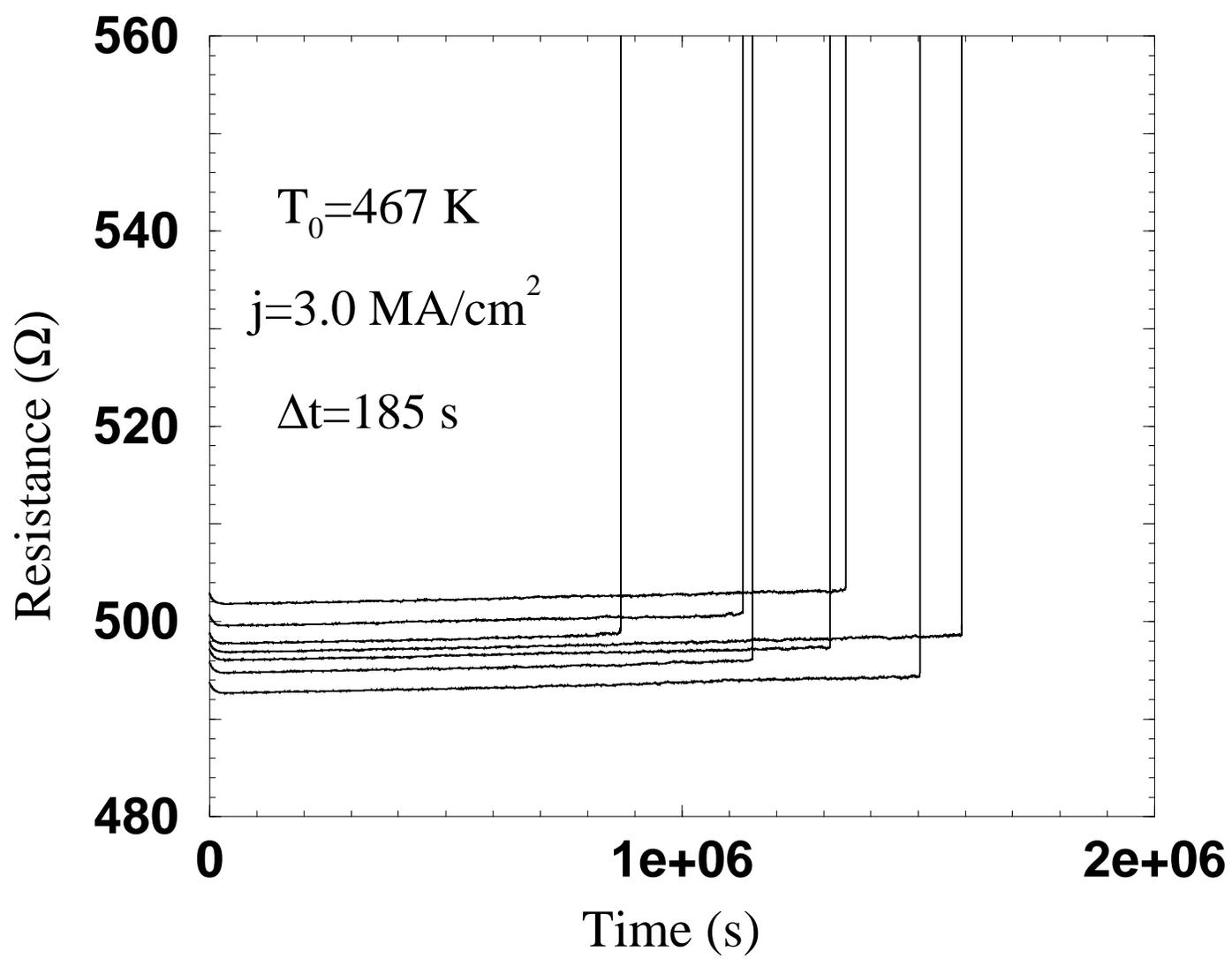
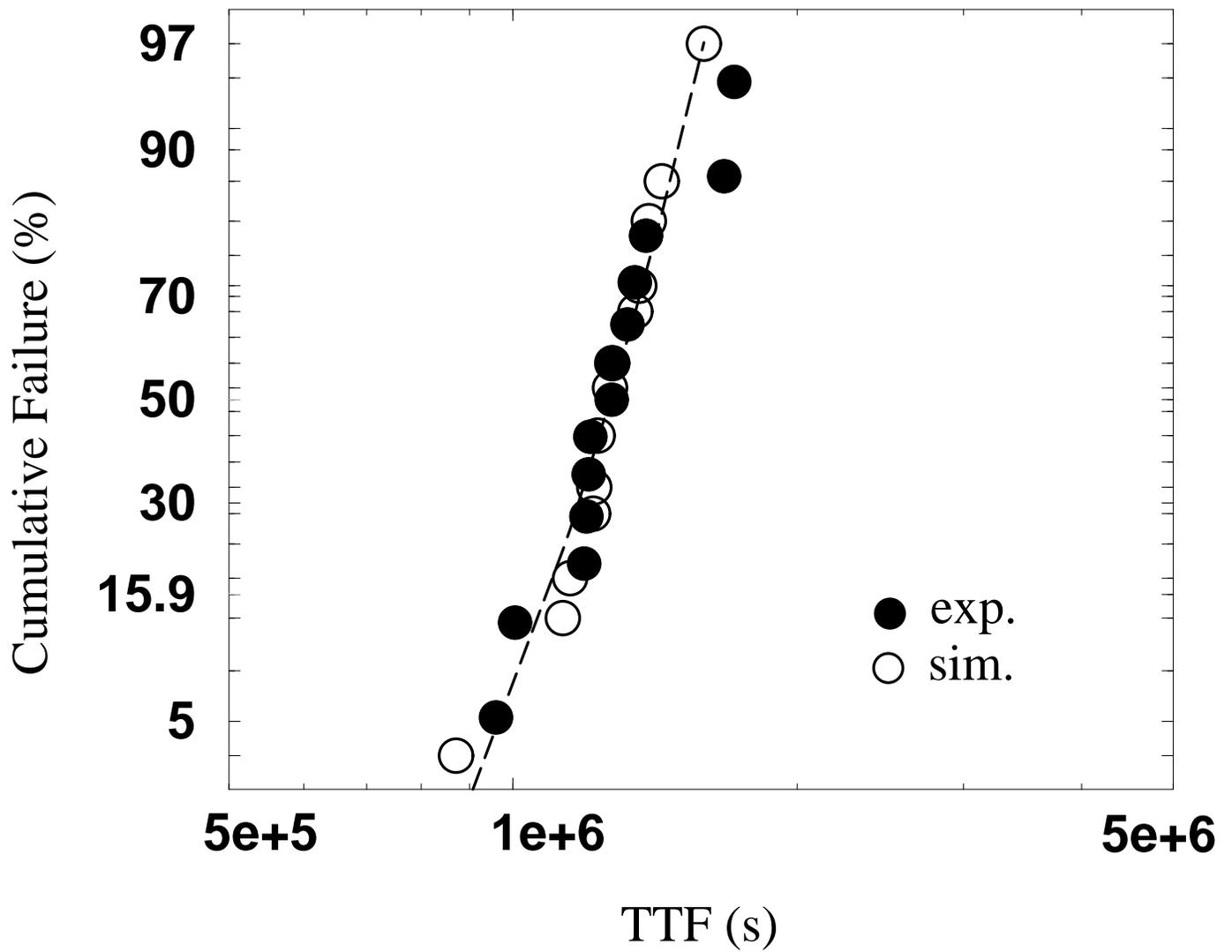
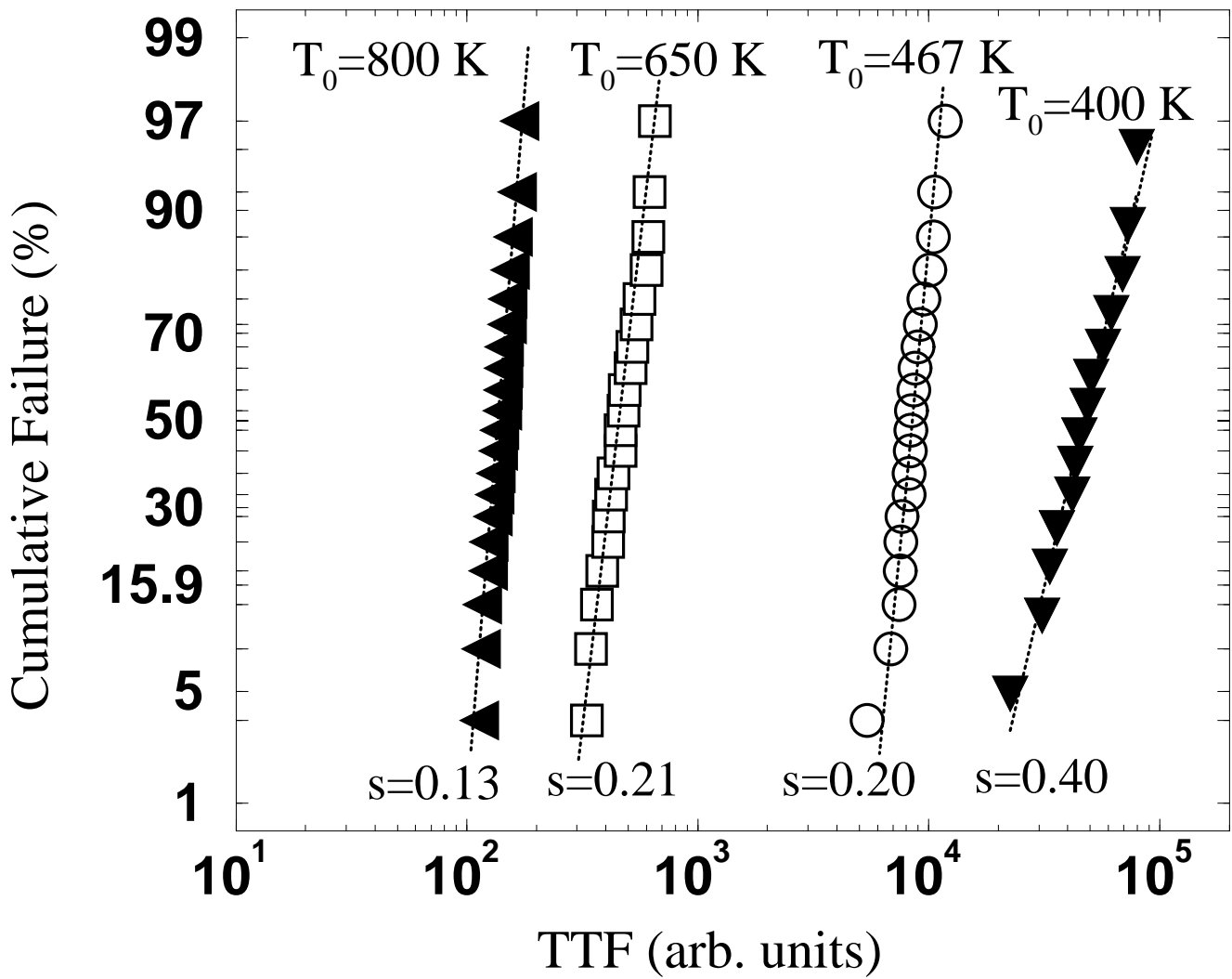
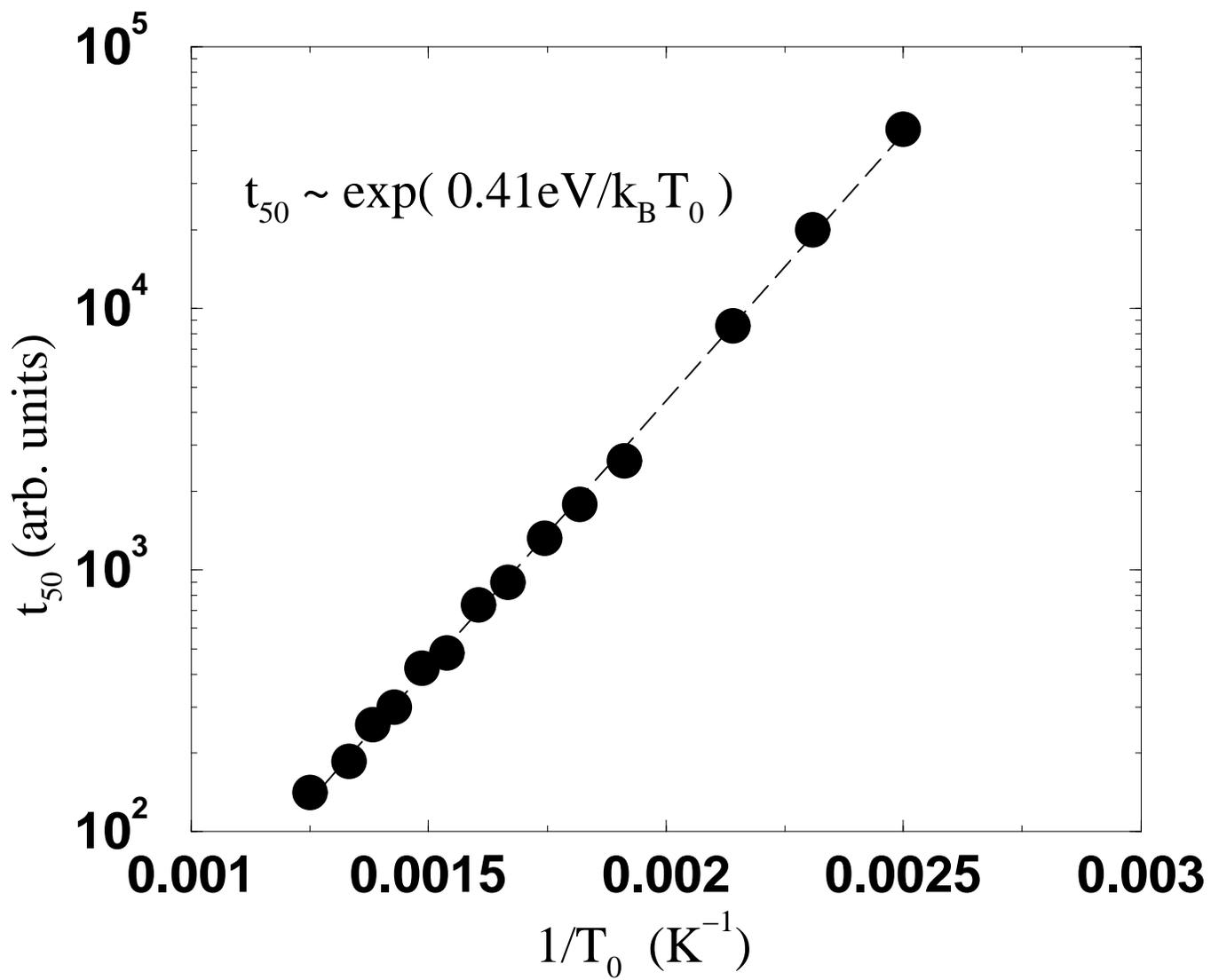
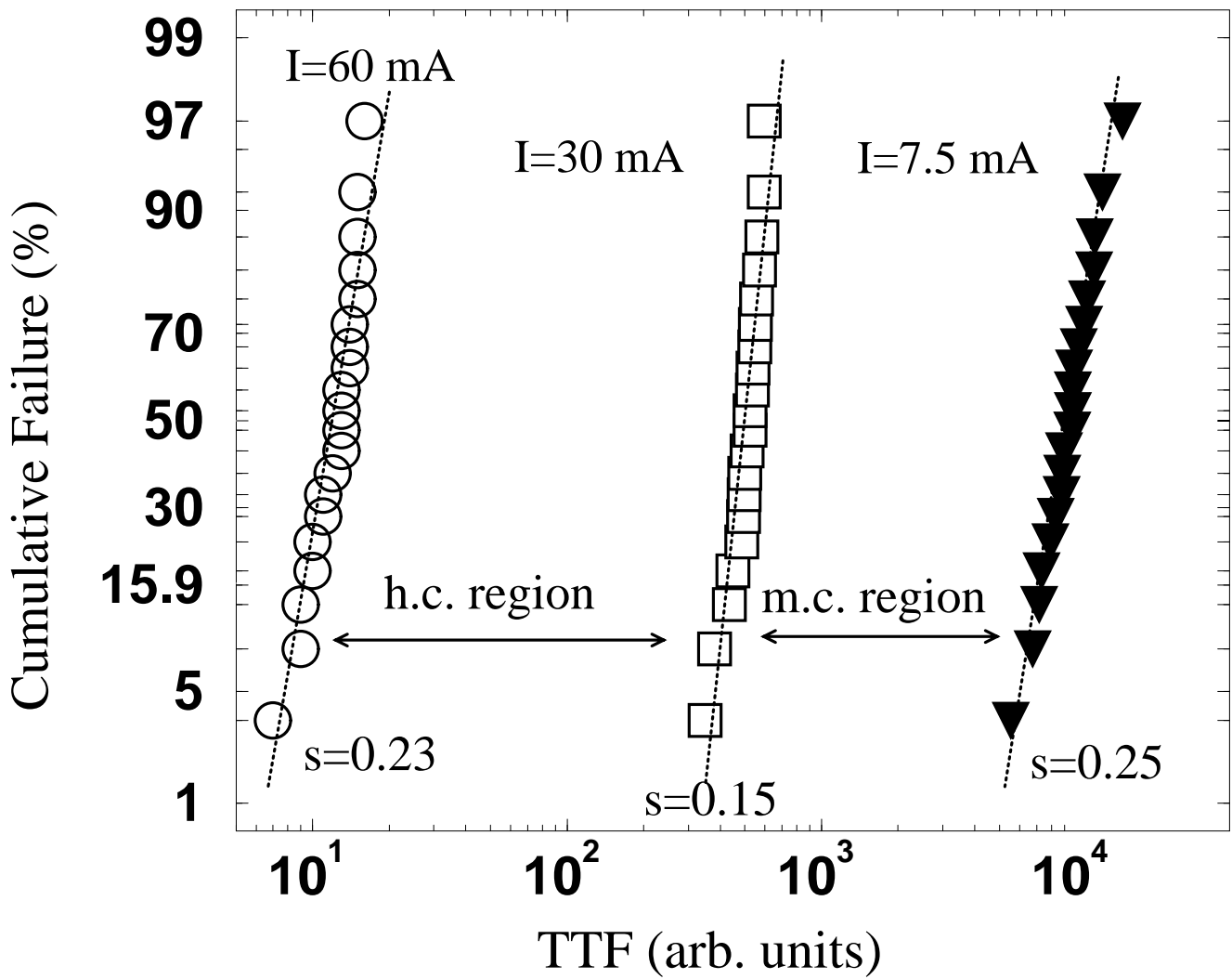
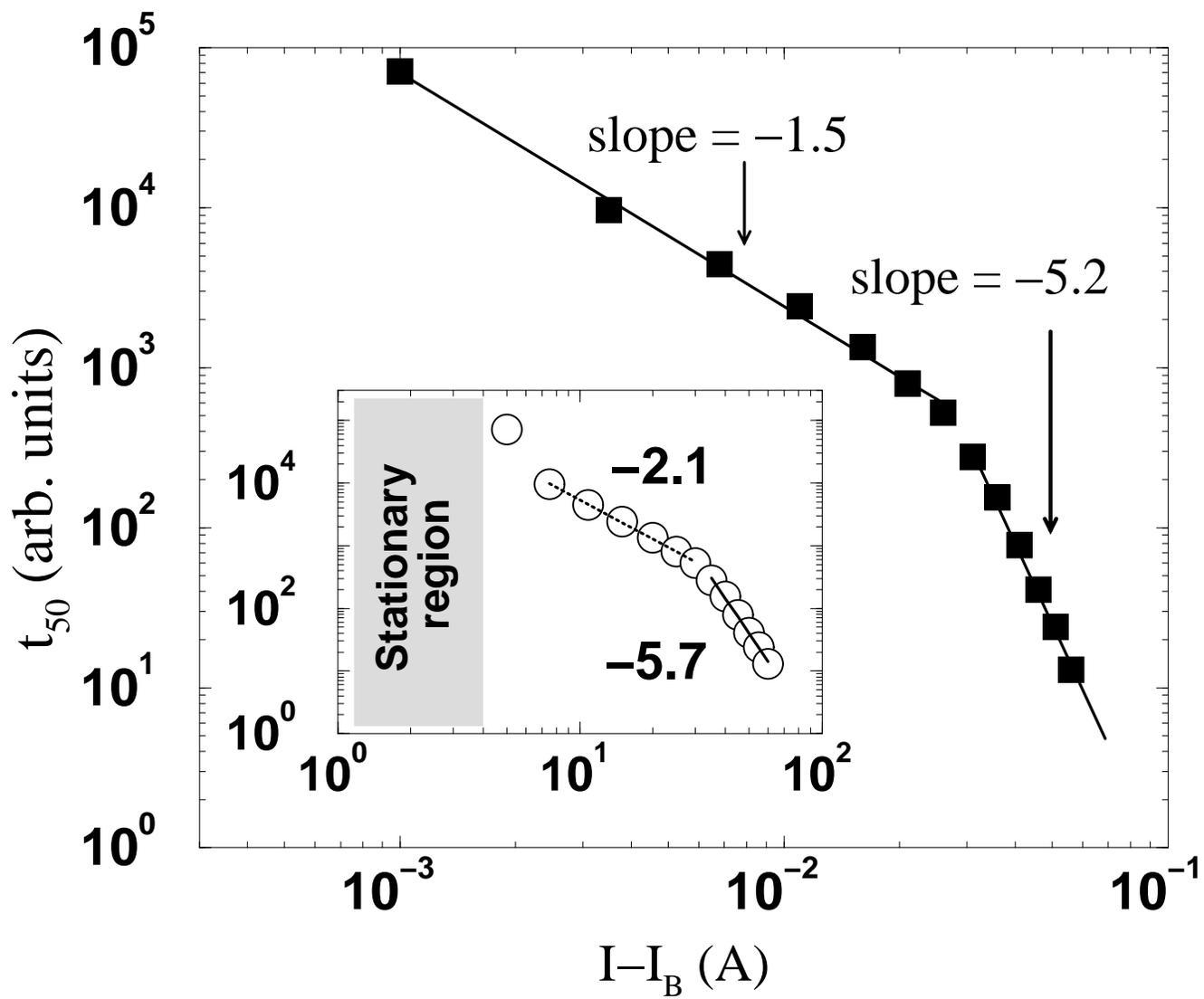
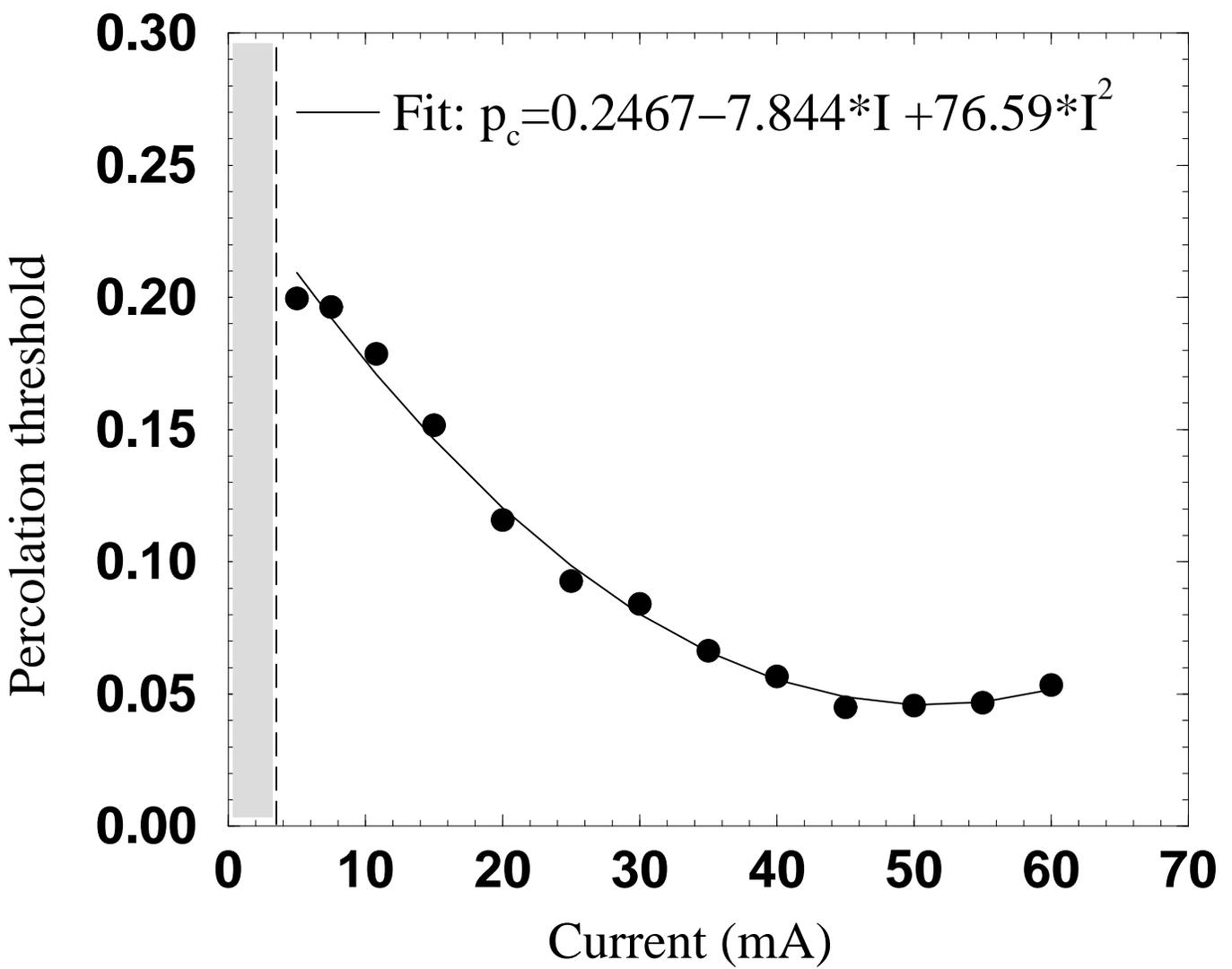
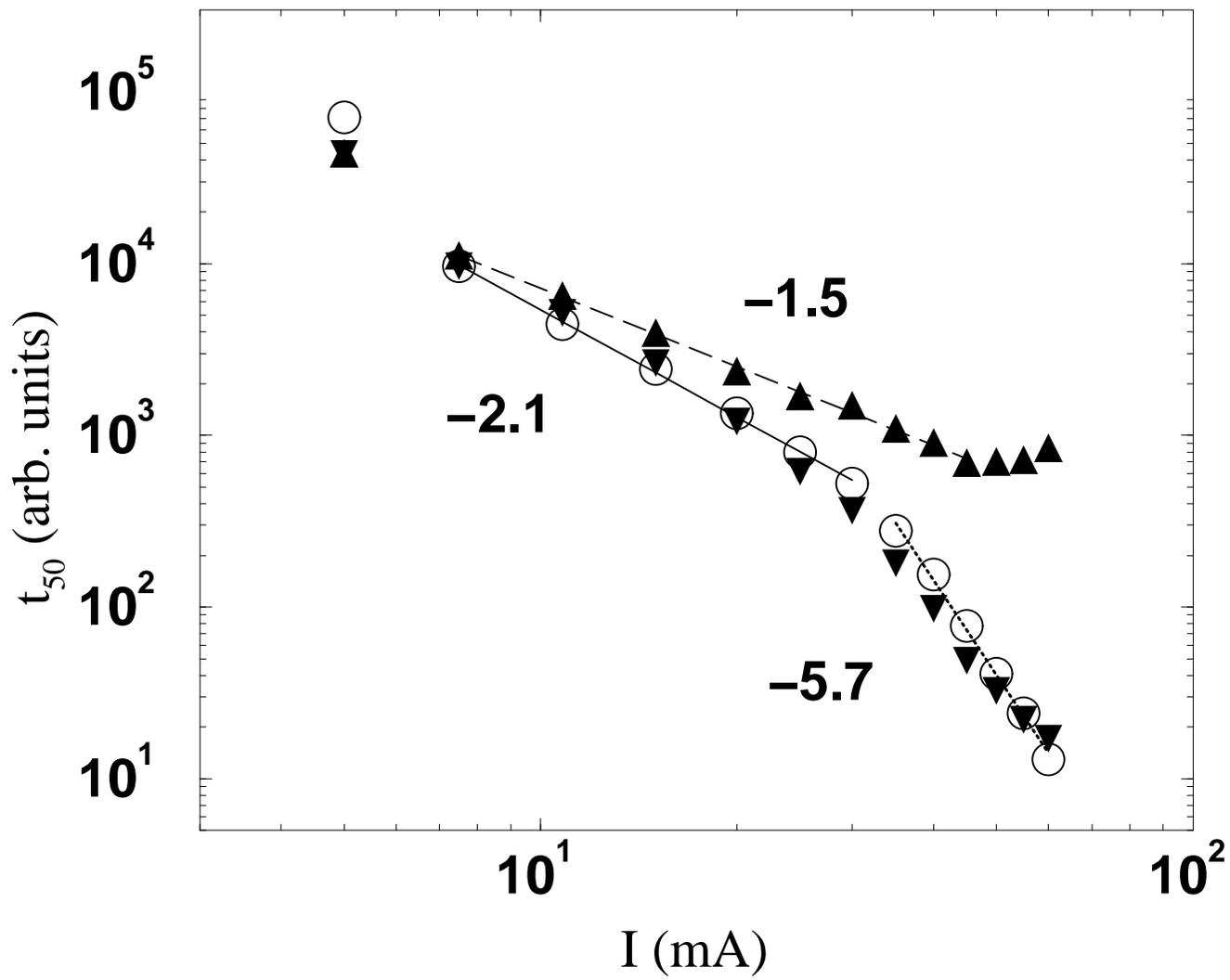
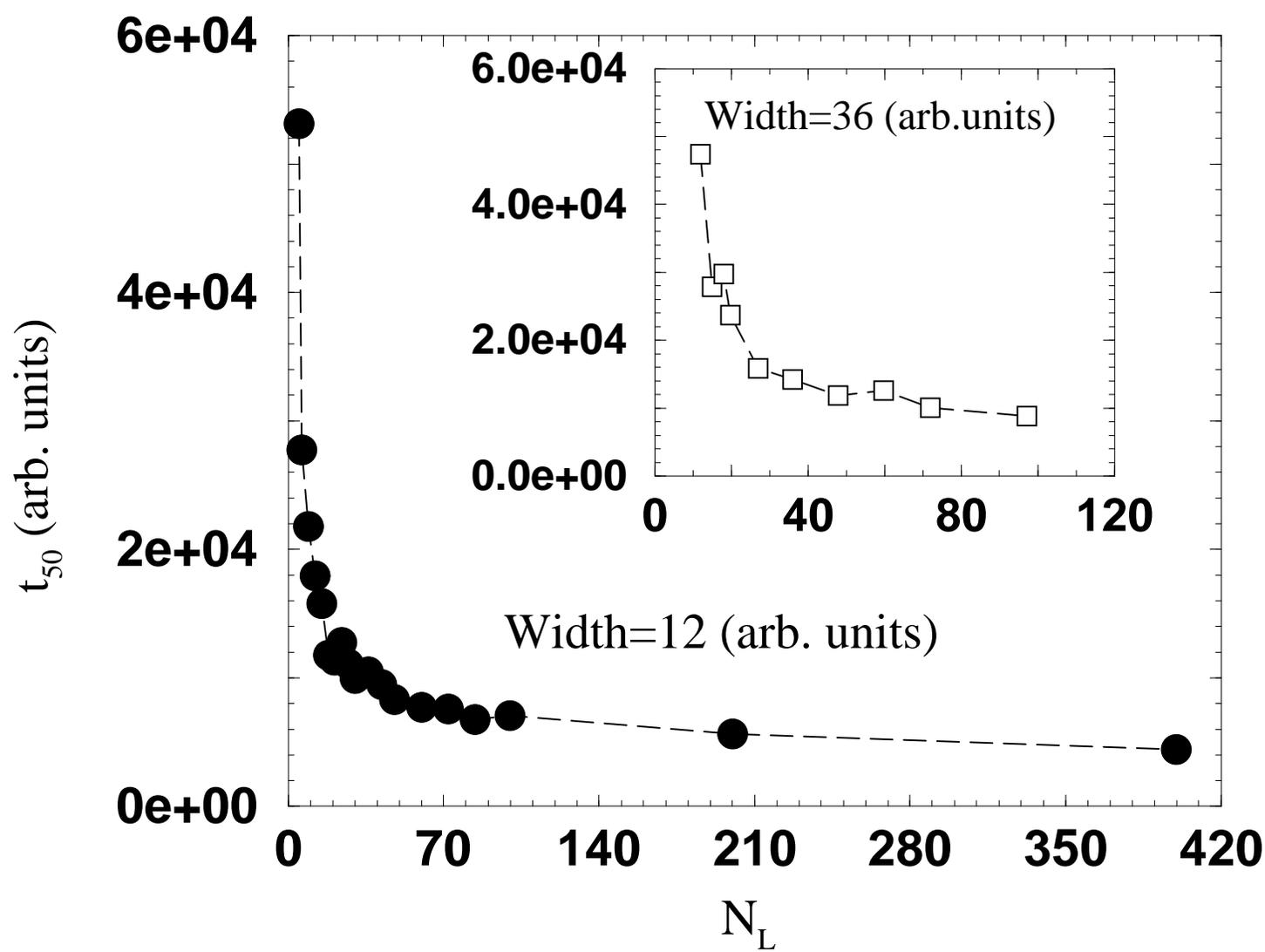
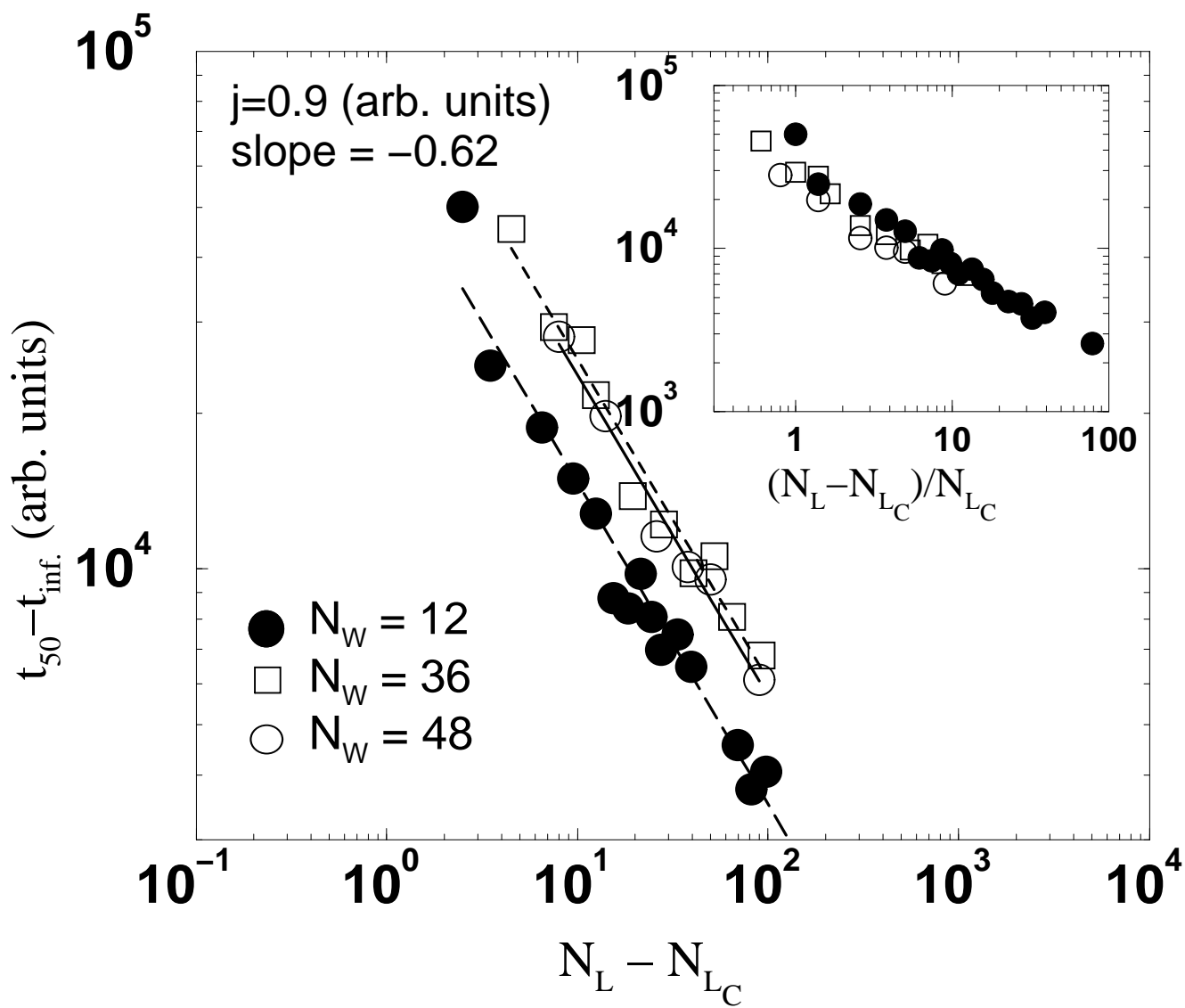
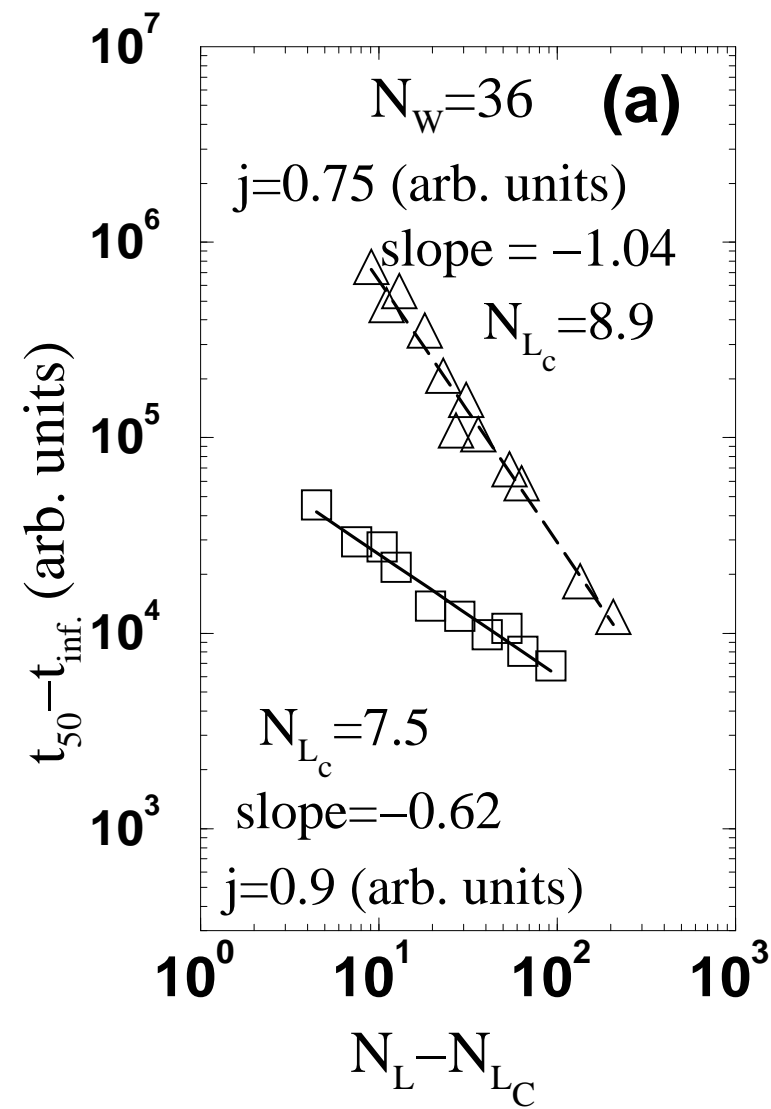
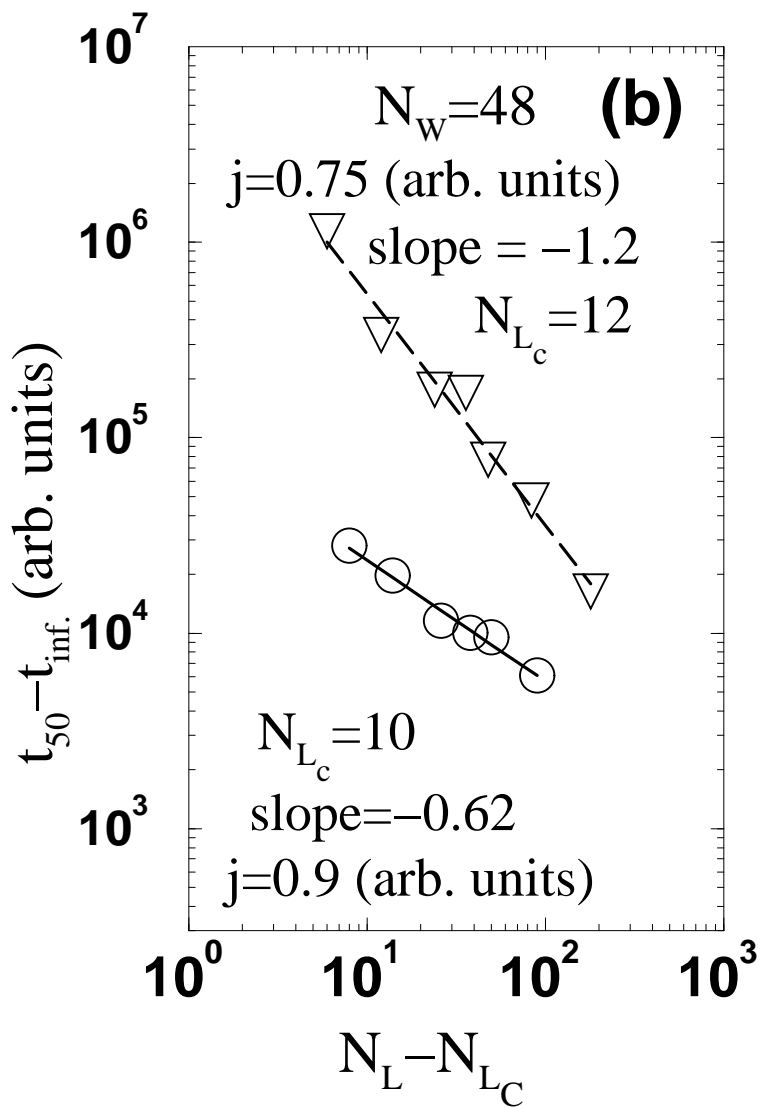
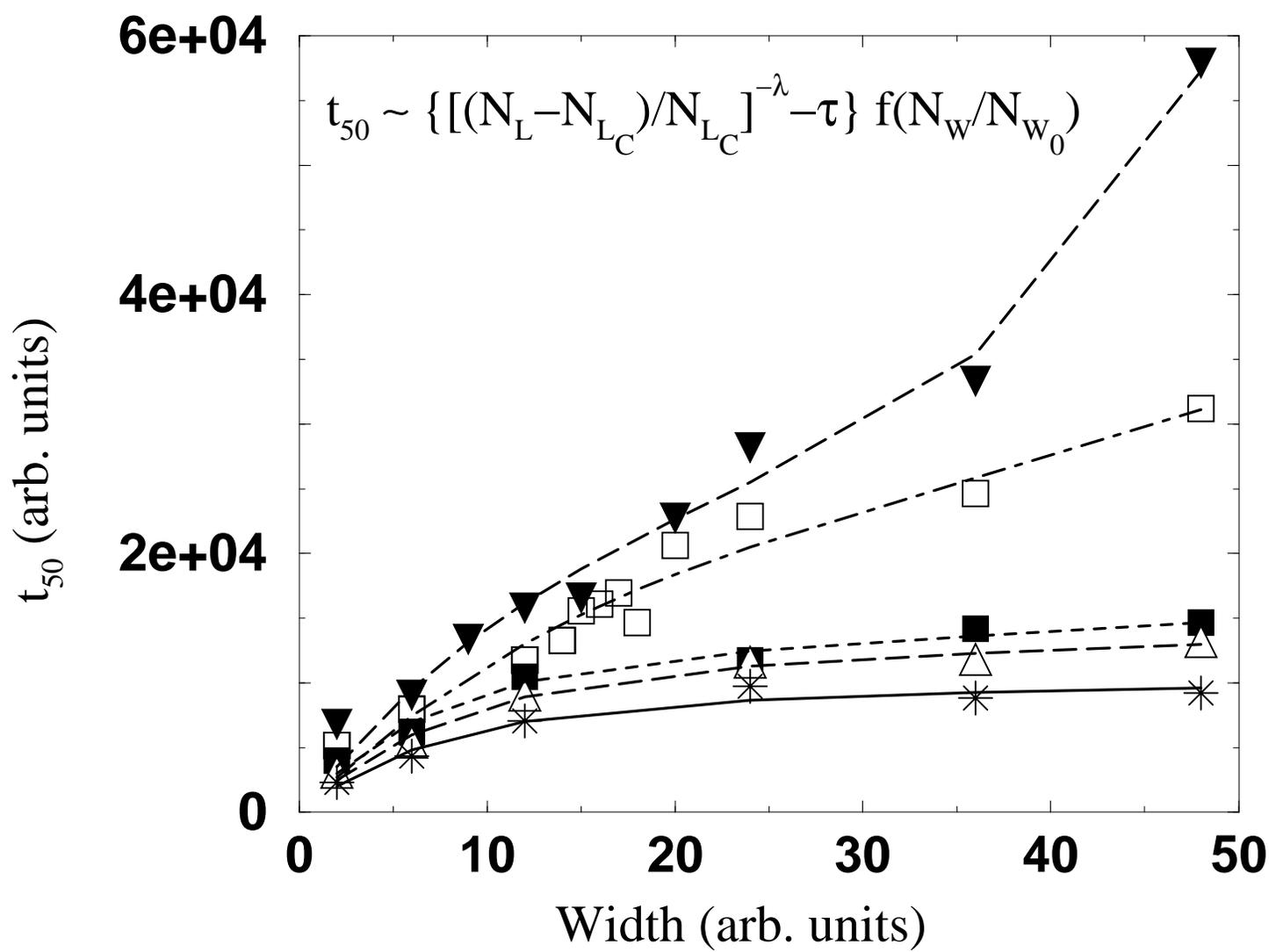
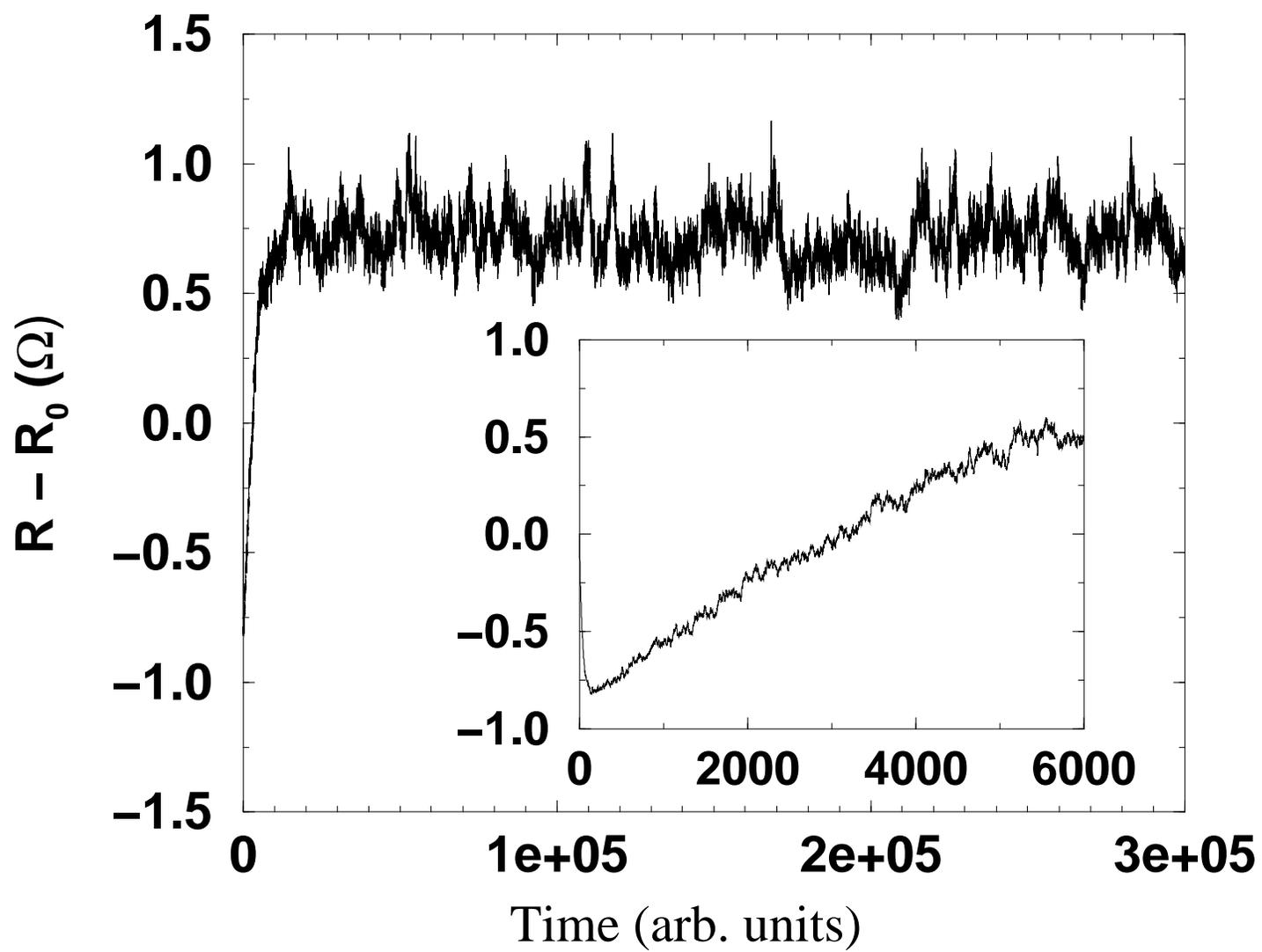
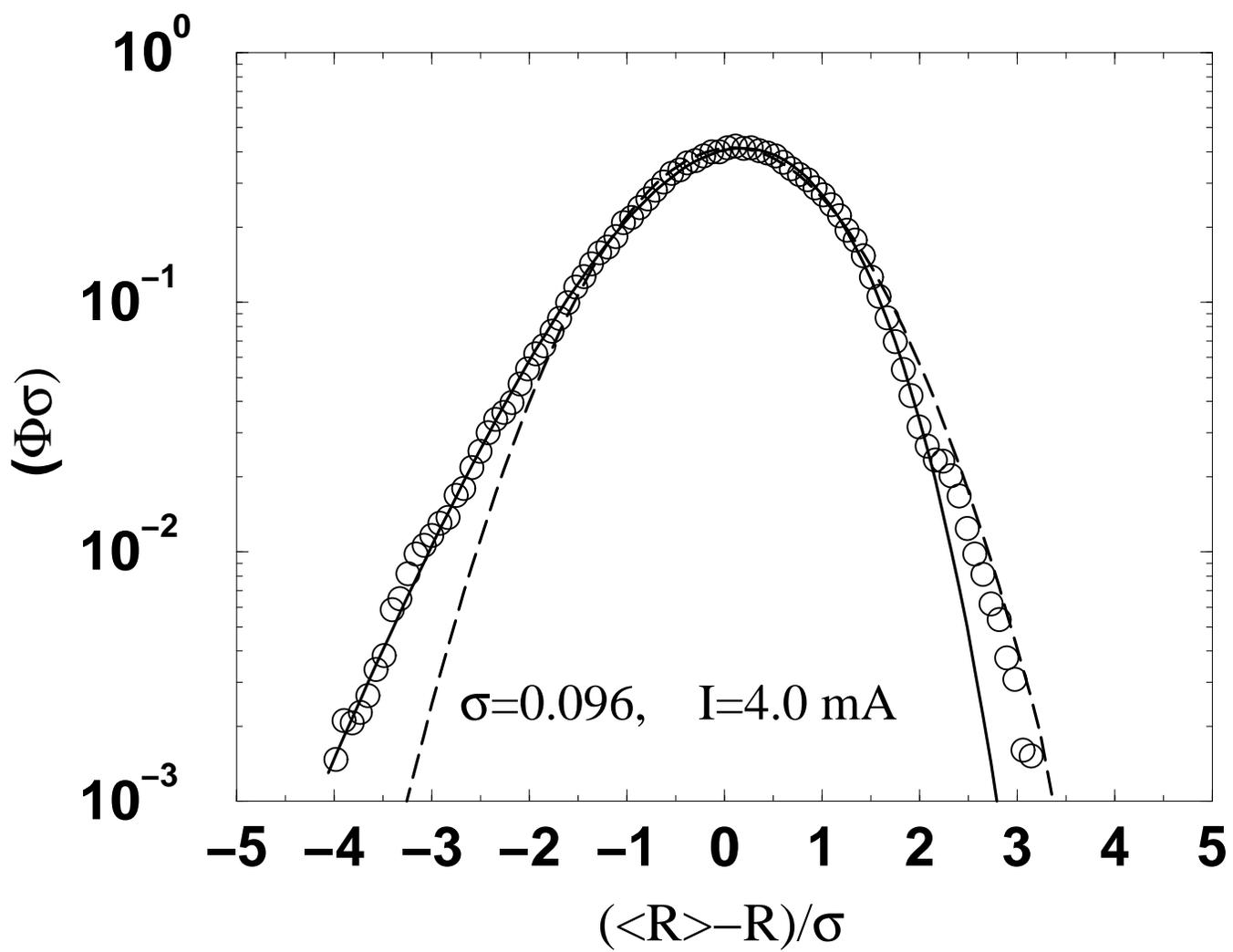
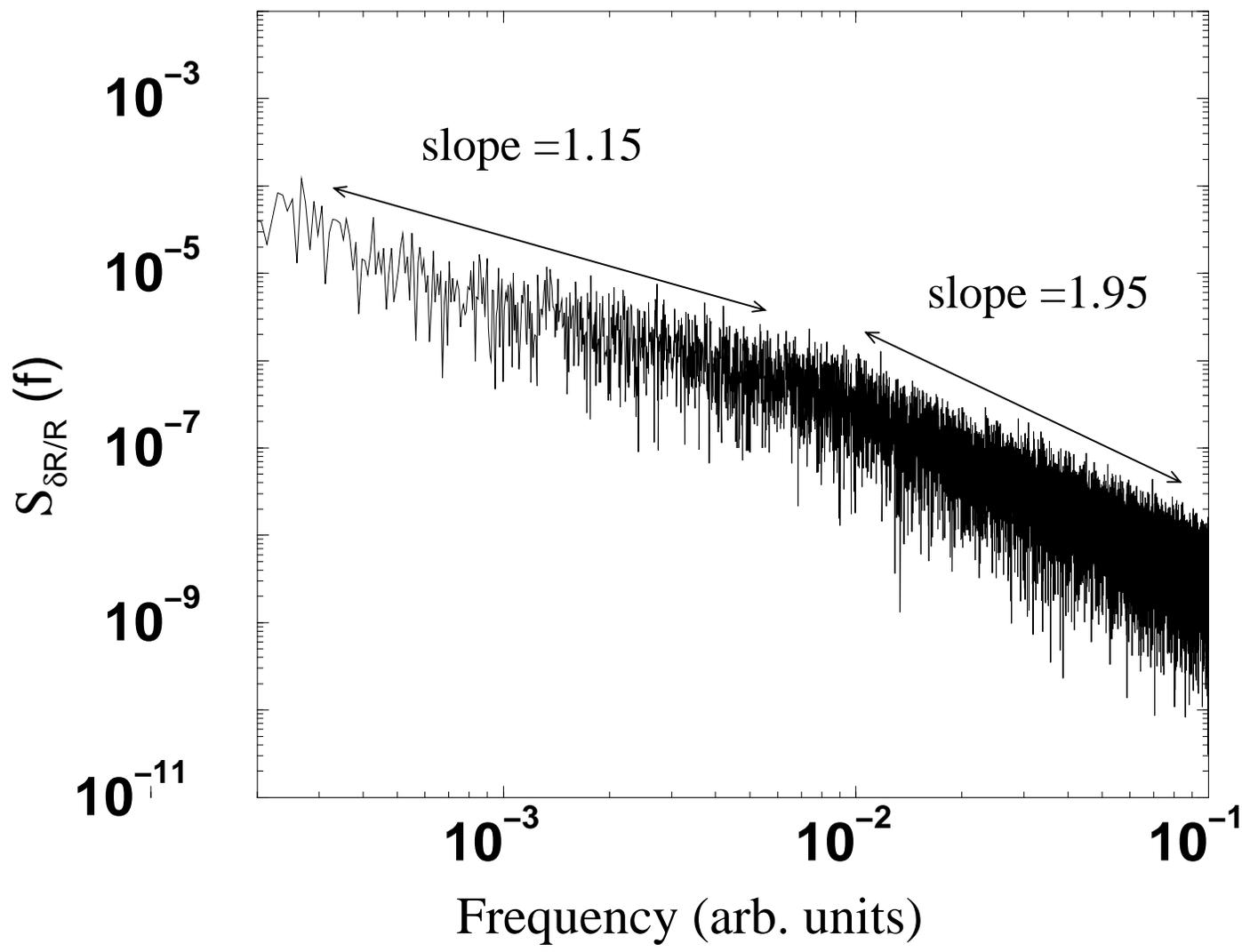
\end{document}